\documentclass[12pt]{article}
\usepackage{graphicx}
\usepackage{amssymb}
\newlength{\extraspace}
\newlength{\extraspaces}
\newcommand{\be}{\begin{equation}
\addtolength{\abovedisplayskip}{\extraspaces}
\addtolength{\belowdisplayskip}{\extraspaces}
\addtolength{\abovedisplayshortskip}{\extraspace}
\addtolength{\belowdisplayshortskip}{\extraspace}}
\newcommand{\ee}{\end{equation}}

\newcommand{\ba}{\begin{eqnarray}
\addtolength{\abovedisplayskip}{\extraspaces}
\addtolength{\belowdisplayskip}{\extraspaces}
\addtolength{\abovedisplayshortskip}{\extraspace}
\addtolength{\belowdisplayshortskip}{\extraspace}}
\newcommand{\ea}{\end{eqnarray}}

\newcommand{\newsection}[1]{
\vspace{12mm}
\pagebreak[3]
\addtocounter{section}{1}
\setcounter{equation}{0}
\setcounter{subsection}{0}
\noindent{\bf \thesection. #1}
\nopagebreak
\medskip
\nopagebreak}

\newcommand{\newsubsection}[1]{
\vspace{0.8cm}
\pagebreak[3]
\addtocounter{subsection}{1}
\noindent{\it \thesubsection. #1}
\nopagebreak
\vspace{2mm}
\nopagebreak}

\newcounter{saveeqn}

\newcommand{\dif}{\mathrm{d}}
\newcommand{\me}{\mathrm{e}}

\begin{document}
\addtolength{\baselineskip}{1.5mm}

\thispagestyle{empty}
\begin{flushright}

\end{flushright}
\vbox{}
\vspace{2cm}

\begin{center}
{\LARGE{Rod-structure classification of gravitational instantons with $U(1)\times U(1)$ isometry
        }}\\[16mm]
{Yu Chen~~and~~Edward Teo}
\\[6mm]
{\it Department of Physics,
National University of Singapore, 
Singapore 119260}\\[15mm]

\end{center}
\vspace{2cm}

\centerline{\bf Abstract}
\bigskip
\noindent
The rod-structure formalism has played an important role in the study of black holes in $D=4$ and 5 dimensions with $\mathbb{R} \times U(1)^{D-3}$ isometry. In this paper, we apply this formalism to the study of four-dimensional gravitational instantons with $U(1) \times U(1)$ isometry, which could serve as spatial backgrounds for five-dimensional black holes. We first introduce a stronger version of the rod structure with the rod directions appropriately normalised, and show how the regularity conditions can be read off from it. Requiring the absence of conical and orbifold singularities will in general impose periodicity conditions on the coordinates, and we illustrate this by considering known gravitational instantons in this class. Some previous results regarding certain gravitational instantons are clarified in the process. Finally, we show how the rod-structure formalism is able to provide a classification of gravitational instantons, and speculate on the existence of possible new gravitational instantons.


\newpage

\newsection{Introduction}

Black holes in higher dimensions have been extensively studied in recent years. It is by now clear that they have a much richer phase structure than their four-dimensional counterparts. For example, in five-dimensional asymptotically flat vacuum space-times, the Emparan--Reall black ring \cite{Emparan:2001} with event horizon topology $S^1 \times S^2$ can in certain cases carry the same mass and angular momentum as the Myers--Perry black hole \cite{Myers:1986} with horizon topology $S^3$. This is in contrast to the situation in four dimensions, in which case the mass and angular momentum uniquely determine a solution, which coincides with the Kerr solution, and the only allowed horizon topology is a sphere $S^2$. Black holes in higher-dimensional asymptotically Kaluza--Klein space-times have an even richer phase structure. For more detailed reviews on the phase structure of black holes in higher dimensions, the reader is referred to \cite{Emparan:2008,Obers:2008,Rodriguez:2010} and references therein.

There is a particular class of higher-dimensional black hole solutions that is more tractable to mathematical analysis, namely stationary vacuum black holes in $D$ space-time dimensions $(D\geq 4)$ with non-degenerate horizons, admitting an additional $D-3$ mutually commuting space-like Killing vector fields (with closed orbits). The symmetries corresponding to these Killing vector fields are referred to as ``axial symmetries", even though in the general case for $D>4$, their fixed-point sets are higher-dimensional surfaces rather than a real axis as in $D=4$. For a static black hole space-time with an additional $D-3$ orthogonal space-like Killing vector fields, it turns out that the Einstein equations decouple into two sets. One of them resembles a three-dimensional flat-space Laplace equation, and the solutions correspond to rod-like sources along a line in the three-dimensional space \cite{Emparan:2001b}. This formalism was subsequently generalised to stationary black hole space-times by Harmark et al.~\cite{Harmark:2004,Harmark:2005}.

Thus each black hole solution in this class will have a certain so-called {\it rod structure\/} associated to it, with the rods themselves physically representing either the event horizon or the symmetry axes. Much information can be read off from a given rod structure, for example, the topology of the event horizon and certain asymptotic properties of the space-time.
Recently, there have also been some attempts to use the rod-structure formalism to extend the four-dimensional black hole uniqueness theorems to higher dimensions. By defining a more mathematical version of the rod structure (known as the interval structure) that takes into account the global properties of the space-time,  Hollands and Yazadjiev \cite{Hollands:2007,Hollands:2008} proved certain uniqueness theorems for stationary black holes which are either asymptotically $\mathbb{R}^{D-1,1}$, or asymptotically $\mathbb{R}^{s,1} \times T^{D-s-1}$ where $0<s<D-1$ (see also \cite{Chrusciel:2008} for more aspects of these space-times).

In this and a subsequent \cite{Chen1} paper, we are interested in this special class of solutions in five space-time dimensions, in particular, those whose two space-like Killing vector fields generate closed orbits. So the isometry group of these solutions is $\mathcal{G}=\mathbb{R}\times \mathcal{T}$, where $\mathbb{R}$ corresponds to the flow of time, and $\mathcal{T}=U(1)\times U(1)$ corresponds to the flows of the two space-like Killing vector fields.\footnote{It is a theorem that every compact, connected, two-dimensional Lie group is commutative, and therefore isomorphic to $U(1)\times U(1)$. This makes it clear why the two commuting space-like Killing vector fields with closed orbits generate an (effective) $U(1)\times U(1)$ isometry group action for the space(-times) considered in this paper.} Well-known solutions belonging to this class include the five-dimensional Myers--Perry black hole and the Emparan--Reall black ring.

When the black hole/ring is removed, the resulting background space-time is usually a direct product of four-dimensional flat space and a flat time dimension. Another possible background space-time that has been considered more recently in the literature \cite{Elvang:2005,Gaiotto:2005,Bena:2005} is the direct product of Euclidean self-dual Taub-NUT space \cite{Newman:1963,Hawking:1976} and a flat time dimension. It turns out that four-dimensional flat space and the self-dual Taub-NUT space are but the simplest examples of {\it gravitational instantons\/}, which are defined to be non-singular four-dimensional Euclidean solutions to the Einstein equations \cite{Gibbons:1979c}. They were extensively studied in the late 1970's and early 1980's within the context of Euclidean quantum gravity (see, e.g., \cite{Gibbons:1994}).

In the present context, these gravitational instantons will be Ricci-flat and have a $U(1)\times U(1)$ isometry group. It turns out that many of the known gravitational instantons fall into this class of manifolds.\footnote{It may well be that the $U(1)\times U(1)$ is only a subgroup of a larger isometry group that the gravitational instanton possesses. But for our purpose, the existence of a $U(1)\times U(1)$ isometry subgroup is a sufficient condition.} Besides the above-mentioned two examples, we also have the Euclidean Schwarzschild and Kerr instantons \cite{Gibbons:1976}, the Eguchi--Hanson instanton \cite{Eguchi:1978}, the Taub-bolt \cite{Page:1979} and Kerr-bolt \cite{Gibbons:1979b} instantons, and the multi-Taub-NUT \cite{Hawking:1976} and Gibbons--Hawking \cite{Gibbons:1979} instantons when all the so-called nuts are collinear. For such gravitational instantons, it is possible to define the rod structure as has been done for five-dimensional black holes.

One of the main aims of this paper is to describe a way in which the possible conical and orbifold singularities of the space(-times) can be readily read off from the rod structure. To do so, we first introduce a stronger version of the rod structure than what has previously been used in the literature, namely one in which the rod directions are normalised to have unit (Euclidean) surface gravity. There is an obvious advantage in adopting this normalisation: the condition that there is no conical singularity along a space-like rod requires that the normalised direction of this rod (as a Killing vector field) generates orbits with period $2\pi$. In order to avoid an orbifold singularity at a so-called turning point where two adjacent space-like rods intersect, their normalised rod directions must generate orbits with period $2\pi$ independently. Together, they can be identified as the pair of independent $2\pi$-periodic generators of the $U(1)\times U(1)$ isometry group. Furthermore, they must be related to any other adjacent direction pair of space-like rods by a $GL(2,\mathbb{Z})$ transformation, so that {\it any\/} adjacent direction pair of space-like rods can serve as the pair of independent $2\pi$-periodic generators of the $U(1)\times U(1)$ isometry group. If these conditions are met, the space(-times) are guaranteed to be free of conical and orbifold singularities.

For each of the known gravitational instantons in the class considered here, we will calculate its rod structure and check the conditions for the space to be free of conical and orbifold singularities. In most cases, this is a straightforward calculation which gives the required identifications of the coordinates. For the case of the self-dual Taub-NUT instanton, we will obtain the same identifications that Misner \cite{Misner:1963} famously found by considering two overlapping coordinate systems to eliminate Dirac-type string singularities. However, our approach has the advantage that it can be applied to other gravitational instantons, no matter how complicated, in a systematic and conceptually unified way. For example, we will be able to derive the required identifications for the multi-collinearly-centered Taub-NUT instanton; something which has not always been done correctly in the literature. Unfortunately, we will also find that the Kerr-bolt solution \cite{Gibbons:1979b,Ghezelbash:2007} is not a true gravitational instanton in the sense that it can never be made completely regular; although appropriate identifications of the coordinates will eliminate the conical singularities, there will always be orbifold singularities at the two fixed points of the $U(1)\times U(1)$ isometry.\footnote{The presence of orbifold singularities is sometimes tolerated in modern contexts such as string theory. However, we shall adopt the more traditional viewpoint that regular manifolds should strictly be free of such singularities.}

Our consideration of the various examples will also illustrate the potential as well as limitations of using the rod structure as a way to distinguish different gravitational instantons. On one hand, the stronger version of the rod structure used in this paper is able to distinguish the Eguchi--Hanson and Taub-bolt instantons, whereas the traditional (Harmark) rod structure \cite{Harmark:2004,Harmark:2005} is unable to do so. On the other hand, we will see that our rod structure is unable to say, tell the presence of NUT charge. For example, the self-dual Taub-NUT instanton has the same rod structure as that of four-dimensional flat space (in appropriate coordinates), while the double-centered Taub-NUT instanton has the same rod structure as the Eguchi--Hanson instanton. This is a manifestation of the fact that, in the terminology to be introduced in Sec.~3, the rod structure is unable to distinguish between asymptotically locally flat gravitational instantons, and their asymptotically Euclidean or asymptotically locally Euclidean counterparts.

Despite these limitations, the rod-structure formalism turns out to be a useful way to classify gravitational instantons with $U(1)\times U(1)$ isometry. We will show how imposing the requirement that adjacent pairs of rod directions are related by $GL(2,\mathbb{Z})$ transformations will lead to restrictions on the directions that the various rods can take. For a given number of turning points, this will allow us to list down all possible rod structures that would correspond to regular manifolds without conical or orbifold singularities if appropriate identifications are made. This is done explicitly for the case of two and three turning points. Although some of them can be associated to known gravitational instantons, there is a countable infinity of new rod structures which can not. It is likely that other considerations (e.g., topological constraints or existence of curvature singularities) will rule out many of these new rod structures.  Nevertheless, it is tantalising to wonder if at least some of them would be associated to as yet undiscovered gravitational instantons.

This paper is organised as follows: In Sec.~2, the stronger version of the rod structure with the rod directions appropriately normalised is introduced for five-dimensional black hole space-times, and the regularity conditions of these space-times and their background spaces are discussed. A brief review of some relevant aspects of gravitational instantons is given in Sec.~3. In Sec.~4, the rod structures of various known gravitational instantons with $U(1)\times U(1)$ isometry are analyzed, and the regularity conditions are checked for each case. Some previous results regarding certain of these gravitational instantons are clarified in the process. In Sec.~5, we show how the regularity conditions can in principle be used to determine all allowed rod structures that could be associated to gravitational instantons, and speculate on the existence of possible new gravitational instantons. The paper ends with a discussion of some open questions and possible extensions of this work.

\newsection{The rod structure and regularity conditions}

\newsubsection{The rod structure}

Consider five-dimensional stationary black hole space-times as solutions to the vacuum Einstein equations. We assume, in addition to the Killing vector field corresponding to time flow $V_{(0)}=\frac{\partial}{\partial t}$, the existence of two linearly independent, commuting, space-like Killing vector fields $V_{(1)}$ and $V_{(2)}$, which also commute with $V_{(0)}$.\footnote{At this point, we do not assume the Killing vector field $V_{(1)}$ or $V_{(2)}$ generates closed orbits. It will be clear below that although $V_{(1)}$ and $V_{(2)}$ together generate a $U(1)\times U(1)$ isometry group, they may not necessarily be the two independent $2\pi$-periodic generators, but instead  may be some linear combinations of them, so in general the orbits of $V_{(1)}$ or $V_{(2)}$ are not periodic.} We also assume the following three assumptions hold (here and henceforth in this subsection, we denote $i,j=0,1,2$):

(1) The tensor $V^{[\mu_0}_{(0)}V^{\mu_1}_{(1)} V^{\mu_{2}}_{(2)} \nabla^\nu V^{\rho ]}_{(i)}$ vanishes at at least one point of the space-time for each $i$.

(2) The tensor $V^{\nu}_{(i)}R_{\nu}{}^{[\rho}V^{\mu_0}_{(0)}V^{\mu_1}_{(1)} V^{\mu_{2}]}_{(2)}=0$ for each $i$.

(3) $\det G$ is non-constant in the space-time, where $G_{ij}=g(V_{(i)},V_{(j)})$ are components of the Gram matrix $G$, and $g$ is the metric of the space-time.

For the Ricci-flat space-times considered in this paper, there exists at least one point where some linear combination of $V_{(1)}$ and $V_{(2)}$ vanishes, so conditions (1)--(3) will be trivially satisfied. Such space-times are referred to as stationary and axisymmetric. It was shown in \cite{Harmark:2004} that for such solutions we can find coordinates $x^i$ (with $x^0=t$), along with $\rho$ and $z$, such that
\be
V_{(i)}=\frac{\partial}{\partial x^i}\,,
\ee
and the metric takes the form
\be
\dif s^2=G_{ij}\dif x^i\dif x^j+\me^{2\nu}(\dif\rho^2+\dif z^2)\,.
\ee
Here $G_{ij}$ and $\nu$ are functions of $\rho$ and $z$ only, and the Gram matrix $G$ is subject to the constraint
\be
\label{rho}
\rho=\sqrt{|\det G|}\,.
\ee
The above coordinates $(x^i,\rho,z)$ are usually referred to as Weyl--Papapetrou coordinates. In these coordinates, the vacuum Einstein equations decouple as
\be
\label{G}
G^{-1}\left(\partial^2_{\rho}+\frac{1}{\rho}\,\partial_{\rho}+\partial^2_{z}\right)G=(G^{-1}\partial_{\rho}G)^2+(G^{-1}\partial_z G)^2,
\ee
and
\ba
\label{nu}
\partial_{\rho}\nu&=&-\frac{1}{2 \rho}+\frac{\rho}{8}\, {\rm Tr}((G^{-1}\partial_{\rho}G)^2-(G^{-1}\partial_z G)^2)\,,\cr
\partial_z\nu&=&\frac{\rho}{4}\,{\rm Tr}(G^{-1}\partial_{\rho}GG^{-1}\partial_z G)\,.
\ea
Notice that the integrability of $\nu$ in (\ref{nu}) is guaranteed by (\ref{rho}) and (\ref{G}). Hence, we can always solve the vacuum Einstein equations by first solving for $G$ using (\ref{rho}) and (\ref{G}), and subsequently solving for $\nu$ using (\ref{nu}).

From the condition (\ref{rho}), it is clear that the Gram matrix is non-degenerate as long as $\rho>0$. At $\rho=0$, it becomes degenerate, so the kernel of $G(\rho=0,z)$ becomes non-trivial, i.e., ${\rm dim}({\rm ker}(G(0,z)))\geq 1$. It was argued in \cite{Harmark:2004} that in order to avoid curvature singularities, it is necessary that ${\rm dim}({\rm ker}(G(0,z)))=1$, except for isolated values of $z$. When this applies, we label these isolated values as $z_1,z_2,\dots,z_N$, with $z_1<z_2<\cdots<z_N$, and call the corresponding points on the $z$-axis ($\rho=0,z=z_i$) {\it turning points\/}. These turning points divide the $z$-axis into $N+1$ intervals $(-\infty,z_1]$, $[z_1,z_2]$,\dots, $[z_{N-1},z_N]$, $[z_N,\infty)$. These intervals are known as {\it rods\/}, assigned to a given stationary and axisymmetric solution. For clarity of presentation, we label these rods from left to right as rod 1, rod 2, \dots, rod $N+1$.

In the interior of a specific rod for $(\rho=0,z_k<z<z_{k+1})$, the Gram matrix has an exactly one-dimensional kernel. It was further shown in \cite{Harmark:2004} that the kernel is constant along the rod. In other words, we can find a constant nonzero vector
\be
v=v^i\frac{\partial}{\partial x^i}=v^i V_{(i)}\,,
\ee
such that
\be
G(0,z)v=0\,,
\ee
for all $z\in [z_k,z_{k+1}]$. The vector $v$ is assigned to this specific rod and is called its direction. For a given solution, the specification of the rods and the directions associated with them is defined as the (Harmark) rod structure of the solution \cite{Harmark:2004,Harmark:2005}. Note that in this definition, the direction of a rod is not unique; it can be any nonzero vector in the one-dimensional kernel of the Gram matrix along the rod.

The direction of a rod defined above is a Killing vector field of the space-time, written in the basis consisting of the three linearly independent and mutually commuting Killing vector fields $V_{(i)}$ of the space-time. Without causing confusion, we sometimes also refer to it as the associated Killing vector field of that rod. Along a specific rod $[z_k,z_{k+1}]$, its associated Killing vector field $v=v^i \frac{\partial}{\partial x^i}$ vanishes. It was shown in \cite{Harmark:2004} that near the interior of the rod $(\rho\rightarrow 0,z_k<z<z_{k+1})$, we have to leading order $g(v,v)=G_{ij}v^i v^j=\pm a(z)\rho^2$ and $\me^{2\nu}=c^2 a(z)$, where $a(z)$ is a function of $z$ and $c$ a constant. Hence, $\frac{G_{ij}v^i v^j}{\rho^2 \me^{2\nu}}$ tends to a constant in the interior of a rod. If it is negative, positive or zero, the rod is said to be time-like, space-like or light-like, respectively.

For the solutions we are interested in, a rod is either time-like or space-like. A time-like rod represents a Killing horizon. If in addition the Killing vector field corresponding to the flow of time is normalised at infinity, i.e., $g(V_{(0)}, V_{(0)})=-1$ at infinity, we can choose a particular direction $v$ in the one-dimensional kernel of the Gram matrix for the horizon rod such that $v=(1,\Omega_1,\Omega_{2})$, where the constants $\Omega_1$ and $\Omega_{2}$ are formally defined to be the angular velocities of the horizon, even though in general, the coordinates $x^1$ and $x^2$ may not correspond to any axes. The surface gravity on the horizon $\kappa={\sqrt{-\frac{1}{2}v_{\mu;\nu}v^{\mu;\nu}}}{\bigg |}_H$ (where $\mu$, $\nu$ run over all the coordinates) is computed to be $\lim\limits_{\rho\rightarrow 0}\sqrt{-\frac{G_{ij}v^i v^j}{\rho^2\me^{2\nu}}}$. Then we can easily see that the Killing vector field $v/\kappa$ has unit surface gravity on the horizon.

If the rod $[z_k,z_{k+1}]$ is space-like, it will represent a (two-dimensional) axis for its associated Killing vector field. Consider the orbits generated by the associated Killing vector field near the interior of the rod $(\rho\rightarrow 0,z_k<z<z_{k+1})$ along a constant $z$ surface. It is easy to see that there will be a conical singularity unless the orbits generated by the associated Killing vector field $v=\frac{\partial}{\partial\eta}$ (the direction of this rod) are identified with period \cite{Harmark:2004}
\be
\Delta \eta=2\pi \lim_{\rho\rightarrow 0}\sqrt{\frac{\rho^2\me^{2\nu}}{G_{ij}v^i v^j}} \,,
\ee
for $z\in (z_k,z_{k+1})$. We define the Euclidean surface gravity on this rod for its associated Killing vector field $v$ as $\kappa_E=\sqrt{\frac{1}{2}v_{\mu;\nu}v^{\mu;\nu}}\bigg |_{\rm rod}=\lim\limits_{\rho\rightarrow 0}\sqrt{\frac{G_{ij}v^i v^j}{\rho^2\me^{2\nu}}}$. Then the Killing vector field $v/\kappa_E$ will have unit Euclidean surface gravity on the rod, and its orbits should be identified with period $2\pi$ in order to avoid a potential conical singularity along the rod.

Thus it is natural to fix the freedom in the direction of a rod by choosing one particular vector in the kernel of the Gram matrix along the interior of the rod, such that it has unit surface gravity for a time-like rod and unit Euclidean surface gravity for a space-like rod.\footnote{This normalisation of the direction was previously performed, e.g., in \cite{Martelli:2005,Cvetic:2005,Lu:2008}.} This fixed direction is referred to as the normalised direction of the rod. For a given solution, the specification of the rods and the normalised directions associated with them is referred to as the {\it rod structure\/} of the solution. From now on, the associated Killing vector field of a rod refers only to its normalised direction, and the rod structure of a solution refers only to the stronger version of the rod structure defined here with the rod directions appropriately normalised.

We note that the normalised direction of a rod defined above can differ by a minus sign, but it does not make a difference in the treatment of this paper. We also note that if the previously defined three linearly independent and mutually commuting Killing vector fields $V_{(i)}$ satisfy conditions (1)--(3), so do the three new Killing vector fields $\tilde{V}_{(i)}=A_{ij}V_{(j)}$ provided the matrix $A_{ij}$ is constant and non-singular. We further take $A_{00}=1$ and $A_{01}=A_{02}=0$, so that $\tilde{V}_{(0)}=V_{{(0)}}$ is also normalised at infinity. Then we can introduce new Weyl--Papapetrou coordinates $({\tilde{x}^i},\tilde{\rho},\tilde{z})$ such that ${\tilde{V}_{(i)}}=\frac{\partial}{\partial {\tilde{x}^i}}$. It can be shown that they are related to the old coordinates $(x^i,\rho,z)$ simply by a linear coordinate transformation
\be
x^i=A_{ji}\tilde{x}^j,\qquad \rho=\frac{1}{|{\rm det}(A_{ij})|}\,\tilde{\rho}\,, \qquad z=\pm\frac{1}{|{\rm det}(A_{ij})|}\,\tilde{z}\,,
\ee
up to harmless translations of $x^i$ and $z$, all of which are chosen to be zero. Also we always have the freedom to choose $z=\frac{1}{|{\rm det}(A_{ij})|}\,\tilde{z}$. So we can clearly see that for a given solution, if $(\rho=0,z=z_k)$ is its $k$-th turning point in the old coordinates, then $(\tilde{\rho}=0,\tilde{z}=|{\rm det}(A_{ij})|\,z_k)$ is its $k$-th turning point in the new coordinates. Furthermore, if $v_k=v_k^i \frac{\partial}{\partial x^i}$ is the normalised direction for its $k$-th rod in the old coordinates, then $\tilde{v}_k=\tilde{v}_k^i \frac{\partial}{\partial \tilde{x}^i}=v_k$ with $\tilde{v}_k^i=v_k^j (A^{-1})_{ji}$ is the normalised direction for its $k$-th rod in the new coordinates. In other words, the normalised directions, though expressed in the new basis consisting of the three Killing vector fields $\tilde{V}_{(i)}$, are invariant under the above coordinate transformation. Hence, for a physical space-time, different choices of the two space-like Killing vector fields, and so the corresponding Weyl--Papapetrou coordinates, lead to slightly different rod structures. However, in most cases, it is advantageous to make a particular choice of the two space-like Killing vector fields, and thus fully determine the Weyl--Papapetrou coordinates and the corresponding rod structure (up to a minus sign for the directions) for a solution. In the following subsection, the rod structure {\it in standard orientation\/} for different cases is defined by making a particular choice of these two space-like Killing vector fields, so that the rod directions have very simple expressions.

Finally, we point out that the results of this subsection are not necessarily confined to five space-time dimensions. They can readily be extended to any $D$-dimensional ($D\geq4$) stationary space-time with $D-3$ linearly independent, mutually commuting, space-like Killing vector fields $V_{(i)}$, $i=1,\dots,D-3$.

\newsubsection{Regularity conditions}

We now focus on the necessary conditions for these black hole space-times to be regular, by which we mean free of conical and orbifold singularities. Let us denote the pair of {\it independent generators\/} of the $U(1)\times U(1)$ isometry group by $\{e_1,e_2\}$, which are assumed to generate the actions of the respective $U(1)$ factors, and are normalised to have period $2\pi$. Then the Killing vector field $\ell=a_1 e_1+a_2 e_2$ will generate a $U(1)$ isometry subgroup whose orbits are periodic with period $2\pi$ if $a_1$ and $a_2$ are coprime integers; on the other hand, if $a_1$ and $a_2$ are any real numbers other than coprime integers, $\ell$ will generate an isometry subgroup with orbits that are either non-periodic, or periodic but with a period different from $2\pi$. Suppose the solution has a space-like rod with (normalised) direction $v$. Recall that to avoid a potential conical singularity, the orbits generated by $v$ must have period $2\pi$ in the vicinity of this rod. So it is clear that we must have $v=a_1 e_1+a_2 e_2$ for some coprime integers $a_1$ and $a_2$.\footnote{We assume that in the basis consisting of $(V_{(0)}=\frac{\partial}{\partial t},e_1,e_2)$, the direction $v$ of any space-like rod does not have a $V_{(0)}$ component. If this were not the case, periodicity conditions imposed on the orbits of $v$ will impose certain identifications on the time coordinate $t$.} Hence, in the basis consisting of $(e_1,  e_2)$, the two components of the direction of any space-like rod must be coprime integers.

If two space-like rods $[z_{k-1},z_k]$ and $[z_k,z_{k+1}]$, with their corresponding directions $v_k$ and $v_{k+1}$ (in the basis $(e_1,  e_2)$), intersect at the turning point $(\rho=0,z=z_k)$, a further condition should be satisfied \cite{Hollands:2007,Hollands:2008}:
\be
\label{condition for absence of orbifold singularity}
\det(v_k^i,v_{k+1}^j)=\pm 1\,,
\ee
for $i,j=1,2$, to avoid a possible orbifold singularity at that point. If this condition is satisfied, the Killing vector fields $\{v_k,v_{k+1}\}$ generate $U(1)$ isometry subgroups with $2\pi$-periodic orbits independently; any Killing vector field that is a linear combination of $v_k$ and $v_{k+1}$ will generate a $U(1)$ isometry subgroup with $2\pi$-periodic orbits if and only if the coefficients are coprime integers. And indeed, since $\{v_k,v_{k+1}\}$ is related to $\{e_1,e_2\}$ by a $GL(2,\mathbb{Z})$ transformation, they can serve as another pair of independent $2\pi$-periodic generators of the $U(1)\times U(1)$ isometry group \cite{Hollands:2007,Hollands:2008}.

Hence, for the regular solutions considered in this paper, we can identify the directions of any two adjacent space-like rods as the pair of independent $2\pi$-periodic generators of the $U(1)\times U(1)$ isometry group. Without loss of generality, we can take the directions of the left-most pair of adjacent space-like rods to be the pair of independent generators of the isometry group. If these two rod directions are related to the directions of the first and last (semi-infinite) rods by a $GL(2,\mathbb{Z})$ transformation, then the latter two rod directions can serve as the pair of independent generators instead. This is not guaranteed to happen, however. Contrast this to the five-dimensional case in \cite{Hollands:2007,Hollands:2008}, where the asymptotic geometry of the space-times considered is the direct product $\mathbb{R}^{s,1} \times T^{5-s-1}$, and the independent generators of the $U(1)\times U(1)$ isometry group are assumed to generate either the standard rotations of the asymptotic Minkowski space-time $\mathbb{R}^{s,1}$ or the flat torus $T^{5-s-1}$. As will be clear below, the solutions considered in this paper have very different and sometimes more complicated asymptotic geometries, from which the pair of independent $2\pi$-periodic generators of the $U(1)\times U(1)$ isometry group cannot be simply identified.

In the case when the first and last (semi-infinite) rods of a particular solution are not parallel, it is useful to introduce Weyl--Papapetrou coordinates defined by taking $\{\tilde{V}_{(1)}=\pm\ell_{N+1},\tilde{V}_{(2)}=\pm\ell_1\}$, where $\ell_1$ and $\ell_{N+1}$ are the (normalised) directions of the first and last rods respectively.\footnote{Of course, we could have taken $\{\tilde{V}_{(1)}=\pm\ell_1,\tilde{V}_{(2)}=\pm\ell_{N+1}\}$ instead. The particular convention above is used so as to be consistent with the examples considered in Sec.~4.} In the event that the first and last rods are parallel, we can introduce Weyl--Papapetrou coordinates defined by taking $\{\tilde{V}_{(1)}=\pm\ell_a,\tilde{V}_{(2)}=\pm\ell_1\}$ instead, where $\ell_a$ is the direction of the second space-like rod from the left such that $\ell_a\neq\pm\ell_1$. In either case, we say that the rod structure has been put in standard orientation. As we shall see in Secs.~4 and 5, putting rod structures in standard orientation is a useful way to check if two rod structures are equivalent up to a coordinate transformation, and to compare different rod structures.

On the other hand, in the particular Weyl--Papapetrou coordinates defined by taking $\{V_{(1)}, V_{(2)}\}$ as the pair of independent $2\pi$-periodic generators $\{e_1,e_2\}$ of the $U(1)\times U(1)$ isometry group, the lengths of the rods in the rod structure will be invariants for isometric space-times. Together with the directions of space-like rods, and certain asymptotic quantities, they were used to characterise a solution in \cite{Hollands:2007,Hollands:2008} for the black hole space-times defined therein. We note that the topology of the event horizon can also be read off from the rod structure; in particular, it is determined solely by the directions of the two space-like rods that are adjacent to the time-like rod representing the horizon.

Even if the solution under consideration is regular, we should point out that Weyl--Papapetrou coordinates are not able to furnish a coordinate chart along the rods. In particular, the metric in these coordinates fails to be analytic at a turning point $(\rho=0,z=z_k)$ where two space-like rods intersect, and local coordinates need to be constructed.
We note that, by identifying the orbits generated by $\{\ell_{k},\ell_{k+1}\}$ with period $2\pi$ independently, the metric in the vicinity of the turning point for a constant time slice can be brought into the standard form of four-dimensional flat Euclidean space $E^4$ near the origin \cite{Dowker:1995,Hollands:2007,Chrusciel:2008}:
\be
\label{R4 origin}
\dif s^2=\dif r_1^2+r_1^2 \dif\phi_1^2+\dif r_2^2+r_2^2 \dif\phi_2^2\,,
\ee
with $r_1,r_2\geq 0$, and with $\frac{\partial}{\partial \phi_1}$ and $\frac{\partial}{\partial \phi_2}$ identified with $\ell_k$ and $\ell_{k+1}$ respectively (so that $\phi_1$ and $\phi_2$ have period $2\pi$ independently).
For more detailed aspects of the behaviour of the space-time near the rods and turning points, and the construction of local coordinates at these locations, the reader is referred to \cite{Hollands:2007,Hollands:2008,Chrusciel:2008,Dowker:1995}.

It may also be worthwhile to point out a fibre-bundle viewpoint of the space-times considered here, following the approach of \cite{Hollands:2007,Hollands:2008,Chrusciel:2008}. Let $M$ be the domain of outer communication of the five-dimensional black hole space-time. Since we assume a $\mathcal{G}=\mathbb{R}\times \mathcal{T}$ isometry group for $M$, it is interesting to see what the orbit space $\hat{M}=M/\mathcal{G}$ is. For the black hole space-times defined in \cite{Hollands:2007,Hollands:2008,Chrusciel:2008}, the orbit space $\hat{M}$ is a two-dimensional manifold with boundaries and corners. It turns out that the same thing holds for the space-times considered here.\footnote{It was assumed in \cite{Hollands:2007}, and subsequently proved in \cite{Hollands:2008}, that the orbit space $\hat{M}$ does not contain conical singularities due to the possible presence of points with discrete isotropy group. We believe that a similar result will hold for our case under suitable technical assumptions, but a proof of this is beyond the scope of the present paper.} We can further map the orbit space analytically to the upper-half complex plane, and introduce globally defined coordinates $(\rho,z)$ such that $\hat{M}=\{z+i\rho \in \mathbb{C}~|~ \rho>0 \}$. These coordinates coincide with the Weyl--Papapetrou coordinates defined above by taking $\{V_{(1)}=e_1, V_{(2)}=e_2\}$, up to a possible translation and reflection of $z$ (we can always appropriately choose them so that they are identical). The boundary $\rho=0$ of $\hat{M}$ consists of a sequence of line segments and corners (intersections of the line segments). The line segments, which correspond to axes or horizons, coincide with the previously defined rods of the solution; while the corners, which correspond to points where the axes intersect, or to points where axes intersect with horizons, coincide with the previously defined turning points. Thus the space-times considered here, with the axes and horizons removed, can be taken as a $\mathcal{G}$-principal fibre bundle over the upper-half complex plane $\hat{M}$, with the projection map naturally sending a point in the space-time to the corresponding point in the orbit space $\hat{M}$.

When we remove the black hole, together with the time dimension $t$, from the space-time (so the Killing vector field $V_{(0)}=\frac{\partial}{\partial t}$ no longer exists), our analysis applies to four-dimensional manifolds $I$ with Euclidean signature, and with an isometry group $\mathcal{T}=U(1)\times U(1)$. This is actually the situation relevant for the gravitational instantons considered in this paper. In this case, any rod in the rod structure is space-like and represents a (two-dimensional) axis, and the turning points are the (zero-dimensional) intersection points of these axes.
The orbit space $\hat{I}=I/\mathcal{T}$ of these manifolds is similar to that of the black hole space-times with isometry group $\mathcal{G}=\mathbb{R}\times \mathcal{T}$ analyzed above, except that there are no boundary segments representing the horizons, and no boundary corners representing points where horizons intersect with axes. With all the axes removed, the isometry group $\mathcal{T}=U(1)\times U(1)$ naturally gives these manifolds a $\mathcal{T}$-principal fibre bundle structure over the base space $\hat{I}=\{z+i\rho \in \mathbb{C}~|~ \rho>0 \}$.

We remark that the space-times or spaces considered in this paper, as manifolds with $\mathbb{R}\times U(1)\times U(1)$ or $U(1)\times U(1)$ action respectively, are uniquely determined by the rod structure \cite{Hollands:2008,Orlik1,Orlik2}. For any given rod structure satisfying the regularity condition (\ref{condition for absence of orbifold singularity}), the manifold can in principle be constructed from it.

Finally, we note that the above necessary regularity conditions can be generalised to stationary black holes in $D>5$ space-time dimensions with $\mathbb{R}\times U(1)^{D-3}$ isometry \cite{Harmark:2004,Hollands:2007,Hollands:2008}, or their regular $(D-1)$-dimensional spatial backgrounds with $U(1)^{D-3}$ isometry. The orbit space will again be a two-dimensional manifold with boundaries and corners homeomorphic to the upper-half complex plane. However, the directions of any two adjacent space-like rods intersecting at a turning point must now satisfy a new condition instead of (\ref{condition for absence of orbifold singularity}), as shown in \cite{Hollands:2008}. This makes the analysis of the necessary regularity conditions at the turning points of these space(-times) more involved than the five-dimensional case considered here.

\newsection{Review of gravitational instantons}

Gravitational instantons are defined as non-singular four-dimensional Euclidean solutions to the Einstein equations \cite{Gibbons:1979c}. As gravitational analogues of Yang--Mills instantons, they are stationary phase points of the path integral in Euclidean quantum gravity, and provide tunnelling amplitudes between topologically distinct gravitational vacua. When the cosmological constant $\Lambda=0$, gravitational instantons are nothing but Ricci-flat Riemannian 4-manifolds. Reviews on gravitational instantons may be found in \cite{Eguchi:1978b,Eguchi:1980,Perry:1980}.

Gravitational instanton symmetries have been classified by Gibbons and Hawking \cite{Gibbons:1979c}. In their paper, one-parameter isometry group actions of a gravitational instanton were classified by their two possible types of fixed points: isolated points called {\it nuts\/} and 2-surfaces called {\it bolts\/}. For example, the isometry parameterised by the Euclidean time coordinate has a bolt at the (Euclidean) horizon for the Euclidean Schwarzschild instanton; while it has two isolated nuts, respectively located at the two poles of the (Euclidean) horizon, for the Euclidean Kerr instanton with nonzero rotation parameter. But often the isometry group of a gravitational instanton is more than one-dimensional, and one has to pick out a particular one-parameter isometry subgroup, to see whether its fixed points are nuts or bolts. If the gravitational instanton admits a $U(1)\times U(1)$ isometry group, there are in fact infinitely many possible choices of this one-parameter isometry subgroup, and the corresponding fixed-point set depends on this choice.

Recall that in the rod structure of a gravitational instanton with $U(1)\times U(1)$ isometry, we have the property that along a (space-like) rod, its associated Killing vector field vanishes. So a rod represents a two-dimensional fixed-point set, and thus a bolt, for its associated Killing vector field. At a turning point where two adjacent rods intersect, both the Killing vector fields associated to the two rods vanish. Thus a turning point is a fixed point for the whole $U(1)\times U(1)$ isometry group. It represents a nut for any Killing vector field that generates isometries in the $U(1)\times U(1)$ group, provided it is linearly independent with each of the directions of the two adjacent rods. Moreover, there are no nuts or bolts away from the rods and turning points, with $\rho>0$. Hence, it will be clear where the nut and bolt fixed points will be for any Killing vector field which generates isometries in the $U(1)\times U(1)$ group, once we know the rod structure of the gravitational instanton.

As mentioned in Sec.~2.2, the gravitational instantons, as manifolds, are uniquely determined by the rod structure, so it is natural to relate their topological invariants to their rod structures. It turns out that the Euler number of these gravitational instantons can be easily read off from the rod structure. The Euler number of a compact manifold $M$ with a one-parameter isometry group is $\chi[M]=\sum\limits_{\rm bolts} \chi_i+\sum\limits_{\rm nuts} 1$, where the bolts and nuts are all referred with respect to that one-parameter isometry group, and $\chi_i$ is the Euler number for the $i$-th bolt \cite{Gibbons:1979c,Gibbons:1979b}. The result does not depend on choice of the one-parameter isometry group. This formula also holds for manifolds with boundary provided that the Killing vector field corresponding to the one-parameter isometry group is either everywhere tangential or is everywhere transverse to the boundary. The gravitational instantons considered in this paper satisfy this condition. For a compact bolt in the current context, it is easy to show that it will always have topology $S^2$, and so have Euler number 2. Then we can see that the Euler number for a gravitational instanton with $U(1)\times U(1)$ isometry is nothing but {\it the number of turning points in the rod structure\/}.

We note that it may also be possible to read off the Hirzebruch signature $\tau[M]$ from the rod structure, using the results of \cite{Gibbons:1979c,Gibbons:1979b}. These results rely upon defining the type $(p,q)$ of a nut for a one-parameter isometry group as was done in \cite{Gibbons:1979c}. Now, it is possible to calculate the $(p,q)$-type of each nut from the rod structure.\footnote{This can be done by constructing consistently oriented coordinate charts near the rods and turning points, and requiring appropriate relations between these local coordinate charts and the directions of rods. By doing this, it is actually possible to fix the directions of all the rods up to an overall sign.} However, there are certain subtleties involving the boundary terms in the formula for $\tau[M]$, and we will not discuss the computation of the Hirzebruch signature nor the $(p,q)$-nut-type formalism any further here.

All the explicitly known gravitational instantons with $\Lambda=0$, except for $T^4$ with a flat metric, are non-compact with ``infinities''. It is convenient to regard these gravitational instantons as compact spaces with boundary, where the boundary recedes to infinity. Depending on the behaviour near their infinities, the known gravitational instantons fall into four types: asymptotically (locally) Euclidean or asymptotically (locally) flat \cite{Gibbons:1979gd}. A gravitational instanton is said to be asymptotically locally Euclidean (ALE) if near infinity it is diffeomorphic to $\mathbb{R}\times (S^3/\Gamma)$ where $\Gamma$ is a non-trivial discrete subgroup of $SO(4)$ with free action on $S^3$, and if the metric tends to the standard flat metric at least as fast as (proper distance)$^{-2}$. In the trivial case when $\Gamma$ is the identity, the gravitational instanton is said to be asymptotically Euclidean (AE). If near infinity, the metric instead tends to $\dif s^2=\dif r^2+r^2(\sigma_1^2+\sigma_2^2)+\sigma_3^2$ at least as fast as (proper distance)$^{-1}$, where $\{\sigma_1,\sigma_2,\sigma_3\}$ are the left-invariant one-forms on $S^3$, the gravitational instanton is said to be asymptotically locally flat (ALF). If near infinity a gravitational instanton is diffeomorphic to $\mathbb{R}\times \mathbb{R}\times S^2$ with identifications made on one of the $\mathbb{R}$'s (along with a possible translation along the azimuthal coordinate of $S^2$), and if the metric tends to a standard flat metric at least as fast as (proper distance)$^{-1}$, the gravitational instanton is said to be asymptotically flat (AF). We should note, however, that an AF gravitational instanton is not an ALF gravitational instanton with $\Gamma=1$.

The topology of infinity can readily be read off from the rod structure. First of all, we note that there is an induced $U(1)\times U(1)$ isometry group action on the boundary surface at infinity. In the orbit space, infinity is then represented by a curve far away from any of the finite rods, intersecting only with the first and last (semi-infinite) rods. Suppose that the normalised directions of these two rods are $\ell_1=(p_1,q_1)$ and $\ell_{N+1}=(p_2,q_2)$ (expressed in the basis ($e_1,e_2$) consisting of the pair of independent $2\pi$-periodic generators of the $U(1)\times U(1)$ isometry), the topology of infinity is then a lens space $L(q_1p_2-p_1q_2,w_1q_1-w_2p_1)$, where the integers $w_1,w_2$ solve the equation $w_1q_2-w_2p_2=\pm 1$ \cite{Hollands:2007}. As we have shown, we can always take $\{e_1=\ell_2,e_2=\ell_1\}$. If $\ell_{N+1}=a\ell_1+b\ell_2$, we have $(p_1,q_1)=(0,1)$ and $(p_2,q_2)=(b,a)$. The topology of infinity is then $L(b,w_1)$, where $w_1$ solves the equation $aw_1-bw_2=\pm 1$. Since we have now $aw_1 = \pm 1$(mod $b$), the topology of infinity is $L(b,w_1)\cong L(b,a)$.

Notice that in the case of an AF gravitational instanton (with $U(1)\times U(1)$ isometry), its infinity will have topology $L(0,1)\cong S^1\times S^2$. The infinity of an AE gravitational instanton will have topology $L(1,0)\cong S^3$. In the case where $\Gamma$ is the cyclic group $\mathbb{Z}_p$, the infinity of an ALE gravitational instanton will have a general lens-space topology $L(p,q)$, with $p\geq2$; on the other hand, the infinity of an ALF gravitational instanton can have either topology $L(1,0)$, or $L(p,q)$ with $p\geq2$.

\newsection{Rod structures of known gravitational instantons}

In this section, we analyze the rod structures of known gravitational instantons with a $U(1)\times U(1)$ isometry group.
We first try to adopt their metrics in the most commonly used form, written in the coordinate system $(\psi,\phi,r,\theta)$. All these metrics are independent of the two coordinates $(\psi,\phi)$. For each gravitational instanton, we then take the two linearly independent and commuting Killing vector fields as $\{V_{(1)}=\frac{\partial}{\partial \psi}, V_{(2)}=\frac{\partial}{\partial \phi}\}$, and define the corresponding Weyl--Papapetrou coordinates $(x^1=\psi,x^2=\phi,\rho,z)$. The rod structure of the gravitational instanton is then analyzed. As mentioned in Sec.~2.2, all the rods will be space-like, and their directions are written in the form $(a_1,a_2)$ for simplicity, which is, in fact, $a_1 \frac{\partial}{\partial \psi}+a_2 \frac{\partial}{\partial \phi}$. The pair of independent $2\pi$-periodic generators of the $U(1)\times U(1)$ isometry group of the gravitational instanton is then identified. We also introduce new Weyl--Papapetrou coordinates $(\tilde{x}^1=\tilde{\psi},\tilde{x}^2=\tilde{\phi},\tilde{\rho},\tilde{z})$, in which the rod structure of the gravitational instanton has standard orientation. Then the topology of constant $r$ surfaces is studied, and we also point out, in several cases, that the rod structure alone cannot uniquely determine a solution.

\newsubsection{Four-dimensional flat space}

Four-dimensional flat (Euclidean) space has the well-known metric
\be
\label{4D Euclid}
\dif s^{2}=\dif r^{2}+{r}^{2}({\dif\theta}^{2}+\sin  ^{2}  \theta \, {{\dif\psi}}^{2}+ \cos  ^{2}  \theta \,{{\dif\phi}}^{2})\,,
\ee
where $r$ and $\theta$ take the ranges $r\geq 0$ and $0\leq \theta \leq \frac{\pi}{2}$.

The Weyl--Papapetrou coordinates $(\psi,\phi,\rho,z)$ are related to the above coordinates by
\be
\rho=\frac{1}{2}\,{r}^{2}\sin 2\theta \,,\qquad z=\frac{1}{2}\,{r}^{2}\cos 2\theta \,.
\ee
In these coordinates, the rod structure has just a single turning point, at $(\rho=0, z=0)$ or $(r=0)$. It divides the $z$-axis into two rods:

\begin{itemize}
\item Rod 1: a semi-infinite rod located at $(\rho=0, z\leq 0)$ or $(r\geq 0, \theta=\frac{\pi}{2})$, with (normalised) direction $\ell_{1}=(0,1)$.

\item Rod 2: a semi-infinite rod located at $(\rho=0, z\geq 0)$ or $(r\geq 0, \theta=0)$, with (normalised) direction $\ell_{2}=(1,0)$.
\end{itemize}
This rod structure is illustrated in Fig.~1.

In this rod structure, the two semi-infinite rods 1 and 2 intersect at the origin. So the orbits generated by $\{\ell_1,\ell_2\}$ should be identified with period $2\pi$ independently to ensure regularity, i.e.,
\be
\label{identification for 4D Euclid}
(\psi, \phi) \rightarrow (\psi, \phi+2 \pi) \,,\qquad (\psi, \phi) \rightarrow (\psi+2 \pi, \phi) \,.
\ee
Here and henceforth, we assume implicitly that all the identifications are made for fixed $(\rho,z)$. Thus the direction pair $\{\ell_1, \ell_2\}$ is identified as the pair of independent $2\pi$-periodic generators of the $U(1)\times U(1)$ isometry group of four-dimensional flat space.

Surfaces of constant $r$ carry a naturally induced $U(1)\times U(1)$ isometry group. They are represented by constant $r$ curves in the $(\rho,z)$ half-plane of the orbit space. The analysis of their topology simply follows from the analysis done in Sec.~3 for the topology of infinity of a gravitational instanton. It is then obvious that surfaces of constant $r>0$ in this case have topology $S^3$; they shrink down to a single point at $r=0$. It turns out that four-dimensional flat space is the unique AE gravitational instanton \cite{Schoen:1979uj}.

\begin{figure}[t]
\begin{center}
\includegraphics{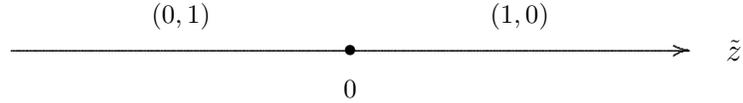}
\caption{The rod structure of four-dimensional flat space and the self-dual Taub-NUT instanton in standard orientation.}
\end{center}
\end{figure}

\newsubsection{Euclidean self-dual Taub-NUT instanton}

The Euclidean self-dual Taub-NUT instanton \cite{Newman:1963,Hawking:1976} has the metric
\be
\label{self-dual Taub-NUT}
\dif s^{2}=H^{-1} { \left( {\dif\psi}+2n\cos  \theta\,
  {\dif\phi} \right) ^{2}}+H ( {\dif r}^{2}+{r}^{2}{
{\dif\theta}}^{2}+{r}^{2} \sin ^2 \theta \,{{\dif\phi}}^{2} )\,,
\ee
where $H$ is a harmonic function on $E^3$ defined as $H=1+\frac{2 |n|}{r}$. The parameter $n$ and coordinates $r$, $\theta$ take the ranges $-\infty <n<\infty$, $r\geq 0$, $0\leq \theta \leq \pi$.

The Weyl--Papapetrou coordinates $(\psi,\phi,\rho,z)$ are related to the above coordinates by
\be
\label{Weyl_TN}
\rho=r\sin \theta  \,,\qquad z=r\cos \theta \,.
\ee
In these coordinates, the rod structure has a single turning point at $(\rho=0, z=0)$ or $(r=0)$, just as that of four-dimensional flat space. It consists of the following two rods:

\begin{itemize}
\item Rod 1: a semi-infinite rod located at $(\rho=0, z\leq 0)$ or $(r\geq 0, \theta=\pi)$, with direction $\ell_{1}=(2 n,1)$.

\item Rod 2: a semi-infinite rod located at $(\rho=0, z\geq 0)$ or $(r\geq 0, \theta=0)$, with direction $\ell_{2}=(-2 n,1)$.
\end{itemize}

In this rod structure, the two semi-infinite rods 1 and 2 intersect at the single turning point. So the orbits generated by $\{\ell_1,\ell_2\}$ should be identified with period $2\pi$ independently to ensure regularity, i.e.,
\be
\label{identification for self-dual Taub-NUT}
(\psi, \phi) \rightarrow (\psi+4 n \pi, \phi+2 \pi) \,,\qquad (\psi, \phi) \rightarrow (\psi-4 n \pi, \phi+2 \pi) \,.
\ee
The direction pair $\{\ell_1, \ell_2\}$ is then identified as the pair of independent $2\pi$-periodic generators of the $U(1)\times U(1)$ isometry group of the self-dual Taub-NUT instanton.

Now, it will prove to be convenient to define a new Killing vector field $\ell_3=\ell_1-\ell_2$. Then $\{\ell_1,\ell_3\}$ is related to $\{\ell_1,\ell_2\}$ by a $GL(2,\mathbb{Z})$ transformation, so $\{\ell_1,\ell_3\}$ can also be taken as the pair of independent $2\pi$-periodic generators of the $U(1)\times U(1)$ isometry group. Since $\ell_3=(4 n,0)$, we thus have another different but equivalent version of the identifications that can give a regular self-dual Taub-NUT instanton:
\be
\label{identification for self-dual Taub-NUT 2}
(\psi, \phi) \rightarrow (\psi+4 n \pi, \phi+2 \pi) \,,\qquad (\psi, \phi) \rightarrow (\psi+8 n \pi, \phi) \,.
\ee

To make contact with other commonly used forms of this gravitational instanton, define a new coordinate $\psi_S=\psi-2 n \phi$ \cite{Misner:1963}. Then the metric (\ref{self-dual Taub-NUT}) takes the form:
\be
\label{self-dual Taub-NUT 2}
\dif s^{2}=H^{-1} { [ {\dif\psi_S}+2n ( 1+ \cos  \theta )\,
  {\dif\phi} ] ^{2}}+H ( {\dif r}^{2}+{r}^{2}{
{\dif\theta}}^{2}+{r}^{2} \sin ^2 \theta\, {{\dif\phi}}^{2} )\,.
\ee
The corresponding Weyl--Papapetrou coordinates are $(\psi_S,\phi,\rho,z)$; by changing to these coordinates, the directions of the two semi-infinite rods are now $\ell_1^\prime=(0,1)$ and $\ell_2^\prime=(-4 n,1)$ respectively (in the basis ($\frac{\partial}{\partial \psi_S}$,$\frac{\partial}{\partial \phi}$)). Then we can take $\{\ell_1^\prime,\ell_1^\prime-\ell_2^\prime\}$ as the pair of independent $2\pi$-periodic generators of the $U(1)\times U(1)$ isometry group of the self-dual Taub-NUT instanton, as they are related to $\{\ell_1^\prime,\ell_2^\prime\}$ by a $GL(2,\mathbb{Z})$ transformation. Since $\ell_1^\prime-\ell_2^\prime=(4 n,0)$, the coordinates $\psi_S$ and $\phi$ in the metric (\ref{self-dual Taub-NUT 2}) should be identified with periods $8 n \pi$ and $2 \pi$ independently, i.e.,
\be
\label{identification for self-dual Taub-NUT 3}
(\psi_S, \phi) \rightarrow (\psi_S, \phi+2 \pi) \,,\qquad (\psi_S, \phi) \rightarrow (\psi_S+8 n \pi, \phi) \,.
\ee
Alternatively, we may define $\psi_N=\psi+2n\phi$ and show that the same identifications as (\ref{identification for self-dual Taub-NUT 3}) (with $\psi_S$ changed to $\psi_N$) should be made. Indeed, the two coordinate systems $(\psi_S,\phi,r,\theta)$ and $(\psi_N,\phi,r,\theta)$ were first used by Misner \cite{Misner:1963} to remove the so-called Dirac--Misner string singularities along the south $(\theta=\pi)$ and north pole $(\theta=0)$ of the gravitational instanton, respectively. To ensure that the two coordinate systems are glued together in a compatible way in their overlapping region, Misner inferred the identifications (\ref{identification for self-dual Taub-NUT 3}) and similarly for $\psi_N$. Here, we have obtained the same identifications by ostensibly different arguments.

We emphasize that the metric (\ref{self-dual Taub-NUT}) together with identifications (\ref{identification for self-dual Taub-NUT 2}), or the metric (\ref{self-dual Taub-NUT 2}) together with identifications (\ref{identification for self-dual Taub-NUT 3}), can give the regular self-dual Taub-NUT instanton. But we note that in the literature identifications such as (\ref{identification for self-dual Taub-NUT 2}) or (\ref{identification for self-dual Taub-NUT 3}) are sometimes misused. Taking for example, the metric (\ref{self-dual Taub-NUT}) with the identifications (\ref{identification for self-dual Taub-NUT 3}) (with $\psi_S$ changed to $\psi$), would result in a space with conical singularities.\footnote{Identifying $\psi$ and $\phi$ in the metric (\ref{self-dual Taub-NUT}) with periods $8n\pi$ and $2\pi$ independently is equivalent to identifying $(4n\frac{\partial}{\partial \psi},\frac{\partial}{\partial \phi})$ as the pair of independent $2\pi$-periodic generators of the $U(1)\times U(1)$ isometry group. In the basis $(4n\frac{\partial}{\partial \psi},\frac{\partial}{\partial \phi})$, we have $\ell_1=(\frac{1}{2},1)$, and $\ell_2=(-\frac{1}{2},1)$. Since the components of $\ell_1$ and $\ell_2$ are now not integer-valued, there are conical singularities along the corresponding two semi-infinite rods. Similar observations have been made in the past by Feinblum \cite{Feinblum}.} In this case, $\phi$ cannot be a periodic coordinate with period $2\pi$, unless when accompanied by a translation in $\psi$.

It is possible to find a new set of coordinates in which the two rod directions have very simple forms. If $n\neq 0$, we have $\ell_1\neq \ell_2$. Then we simply take $\{\tilde{V}_{(1)}=\ell_2,\tilde{V}_{(2)}=\ell_1\}$ and define the corresponding new Weyl--Papapetrou coordinates $(\tilde{x}^1=\tilde{\psi},\tilde{x}^2=\tilde{\phi},\tilde{\rho},\tilde{z})$, such that $\tilde{V}_i=\frac{\partial}{\partial \tilde{x}^i}$. It is easy to show they are related to the old coordinates (\ref{Weyl_TN}) by
\be
\label{coordinate transformation for self-dual Taub-NUT}
\psi=-2 n (\tilde{\psi}-\tilde{\phi})\,,\qquad
\phi=\tilde{\psi}+\tilde{\phi}\,,\qquad
\rho=\frac{1}{4|n|}\tilde{\rho}\,,\qquad
z=\frac{1}{4|n|}\tilde{z}\,.
\ee
In these coordinates, the single turning point is located at $(\tilde{\rho}=0,\tilde{z}=0)$, and the directions of the two semi-infinite rods are simply $K_1=(0,1)$ and $K_2=(1,0)$. This is illustrated in Fig.~1. We say that the rod structure of the self-dual Taub-NUT instanton has standard orientation in the new Weyl--Papapetrou coordinates $(\tilde{\psi},\tilde{\phi},\tilde{\rho},\tilde{z})$. The direction pair $\{K_1,K_2\}$ is then identified as the pair of independent $2\pi$-periodic generators of the $U(1)\times U(1)$ isometry group. So in the new Weyl--Papapetrou coordinates, instead of (\ref{identification for self-dual Taub-NUT}), the following identifications should be made to ensure regularity:
\be
\label{identification for self-dual Taub-NUT 4}
(\tilde{\psi}, \tilde{\phi}) \rightarrow (\tilde{\psi},\tilde{ \phi}+2 \pi) \,,\qquad (\tilde{\psi}, \tilde{\phi}) \rightarrow (\tilde{\psi}+2 \pi, \tilde{\phi}) \,.
\ee

Now we can see that, in appropriately chosen coordinates, the self-dual Taub-NUT instanton with nonzero NUT charge $n$ has exactly the same rod structure as four-dimensional flat space. Furthermore, the self-dual Taub-NUT instanton with different NUT charges $n$ all share the same rod structure. This is an example of the fact that the rod structure alone cannot uniquely determine a solution.

From the rod structure in standard orientation, we can easily see that, just as in the case of four-dimensional flat space, surfaces of constant $r>0$ have topology $S^3$, and they shrink down to a point at $r=0$. At infinity $r\rightarrow \infty$, the self-dual Taub-NUT instanton approaches a finite $S^1$ fibre bundle over an $S^2$. This is nothing but the well-known Hopf fibration of $S^3$. The Killing vector field $\frac{\partial}{\partial \psi}$ generates the finite $S^1$ fibre at infinity, with a constant size $8n\pi$. It can be checked that the self-dual Taub-NUT instanton is ALF with $\Gamma=1$. When $n\rightarrow0$, the $S^1$ dimension vanishes,\footnote{This can be seen from the metric (\ref{self-dual Taub-NUT}), together with the identifications (\ref{identification for self-dual Taub-NUT 2}). $\psi$ has a period which vanishes when $n\rightarrow0$, so the $S^1$ dimension shrinks down to zero.} and we thus recover a three-dimensional flat space; on the other hand, when $n\rightarrow\infty$, the $S^1$ dimension blows up, and we recover a four-dimensional flat space.

We note that the above non-uniqueness result does not violate the uniqueness theorem proved by Hollands and Yazadjiev \cite{Hollands:2008}, as the self-dual Taub-NUT instanton is not within the class of spatial backgrounds of the solutions considered in \cite{Hollands:2008}. Recall that the asymptotic geometry of the space-times considered in \cite{Hollands:2008} is $\mathbb{R}^{s,1}\times T^{D-s-1}$, with an $\mathbb{\mathbb{R}}\times U(1)^{D-3}$ isometry group. The latter generates translations along time, and the standard rotations in the Minkowski space-time $\mathbb{R}^{s,1}$ and the flat torus $T^{D-s-1}$. On the other hand, the self-dual Taub-NUT instanton approaches a non-trivial $S^1$ fibre bundle over $S^2$ at infinity, with a crucial non-vanishing $g_{\psi\phi}$ cross term. In this case, the pair of independent $2\pi$-periodic generators $\{\ell_1,\ell_2\}$ of the $U(1)\times U(1)$ isometry group generates rotations along the south and north poles respectively of a distorted $S^3$.

\newsubsection{Euclidean Schwarzschild instanton}

The Euclidean Schwarzschild instanton has the metric
\be
\label{Euclidean Sch}
{\dif s}^{2}= \left( 1-{\frac {2 m}{r}} \right) {{\dif\psi}}^{2}+\left( 1-{\frac {2 m}{r}} \right) ^{-1}{\dif r}^{2} +{r}^{2} ( {
{\dif\theta}}^{2}+ \sin ^{2} \theta \, {{
\dif\phi}}^{2} )\,,
\ee
where the parameter $m$ and coordinates $r$, $\theta$ take the ranges $r\geq 2 m\geq0$, $0\leq \theta \leq \pi$.

The Weyl--Papapetrou coordinates $(\psi,\phi,\rho,z)$ are related to the above coordinates by
\be
\label{Weyl_Sch}
\rho=\sqrt {{r}^{2}-2m r}\,\sin \theta \,,\qquad z= \left( r-m \right) \cos \theta \,.
\ee
In these coordinates, the rod structure has two turning points at $(\rho=0, z=z_1\equiv-m)$ or $(r=2 m, \theta=\pi)$, and at $(\rho=0, z=z_2\equiv m)$ or $(r=2 m, \theta=0)$. They divide the $z$-axis into three rods; from left to right they are:

\begin{itemize}
\item Rod 1: a semi-infinite rod located at $(\rho=0, z\leq z_1)$ or $(r\geq 2 m, \theta=\pi)$, with direction $\ell_{1}=(0,1)$.

\item Rod 2: a finite rod located at $(\rho=0, z_1\leq z\leq z_2)$ or $(r=2 m, 0\leq\theta\leq\pi)$, with direction $\ell_{2}=(4 m,0)$.

\item Rod 3: a semi-infinite rod located at $(\rho=0, z\geq z_2)$ or $(r\geq 2 m, \theta=0)$, with direction $\ell_{3}=(0,1)$.
\end{itemize}

It is straightforward to check that the direction pairs $\{\ell_1,\ell_2\}$ and $\{\ell_2,\ell_3\}$ of adjacent rods are related by a $GL(2,\mathbb{Z})$ transformation. To ensure regularity, the orbits generated by say the first direction pair $\{\ell_1,\ell_2\}$ should be identified with period $2\pi$ independently, i.e.,
\be
\label{Sch identification}
(\psi, \phi) \rightarrow (\psi, \phi+2 \pi) \,,\qquad (\psi, \phi) \rightarrow (\psi+8 m \pi, \phi)\,.
\ee
The direction pair $\{\ell_1,\ell_2\}$, or equivalently $\{\ell_2,\ell_3\}$, is then identified as the pair of independent $2\pi$-periodic generators of the $U(1)\times U(1)$ isometry group of the Euclidean Schwarzschild instanton.

If $m\neq 0$, we can put the rod structure in standard orientation by taking $\{\tilde{V}_{(1)}=\ell_2,\tilde{V}_{(2)}=\ell_1\}$. The corresponding new Weyl--Papapetrou coordinates $(\tilde{\psi},\tilde{\phi},\tilde{\rho},\tilde{z})$ are related to the old coordinates (\ref{Weyl_Sch}) by
\be
\psi=4 m \tilde{\psi} \,,\qquad
\phi=\tilde{\phi} \,,\qquad
\rho=\frac{1}{4m}\tilde{\rho}\,,\qquad
z=\frac{1}{4m}\tilde{z}\,.
\ee
The two turning points are now pushed to $(\tilde{\rho}=0,\tilde{z}=\tilde{z}_1\equiv-4 m^2)$ and $(\tilde{\rho}=0,\tilde{z}=\tilde{z}_2\equiv4 m^2)$, and the corresponding directions of the three rods from left to right are $K_1=(0,1)$, $K_2=(1,0)$ and $K_3=(0,1)$. This is illustrated in Fig.~2(a). It is clear that in the new Weyl--Papapetrou coordinates, the following identifications should be made to ensure regularity:
\be
(\tilde{\psi}, \tilde{\phi}) \rightarrow (\tilde{\psi},\tilde{ \phi}+2 \pi) \,,\qquad (\tilde{\psi}, \tilde{\phi}) \rightarrow (\tilde{\psi}+2 \pi, \tilde{\phi}) \,.
\ee

The topology of the constant $r>r_0$ surfaces is $S^1\times S^2$, with $\ell_2$ and $\ell_1$ being the generators of the rotations for $S^1$ and $S^2$ respectively. When $r\rightarrow2m$, the $S^1$ vanishes, and the constant $r$ surface becomes a two-sphere $S^2$ at $r=2m$. At infinity $r\rightarrow \infty$, the $S^1$ dimension approaches a constant size $8m\pi$. It can be checked that the Euclidean Schwarzschild instanton is AF. When $m\rightarrow0$, the $S^1$ dimension vanishes, and we recover a three-dimensional flat space.

As a final remark, note that if we take the limit $m\rightarrow0$ without imposing the second identification of (\ref{Sch identification}), we obtain a $U(1)\times\mathbb{R}^3$ space with flat metric, in which the $\psi$ coordinate parameterising the $U(1)$ can have any period. It is the trivial AF gravitational instanton, whose rod structure consists of a single infinite rod with direction (0,1).

\begin{figure}[t]
\begin{center}
\includegraphics{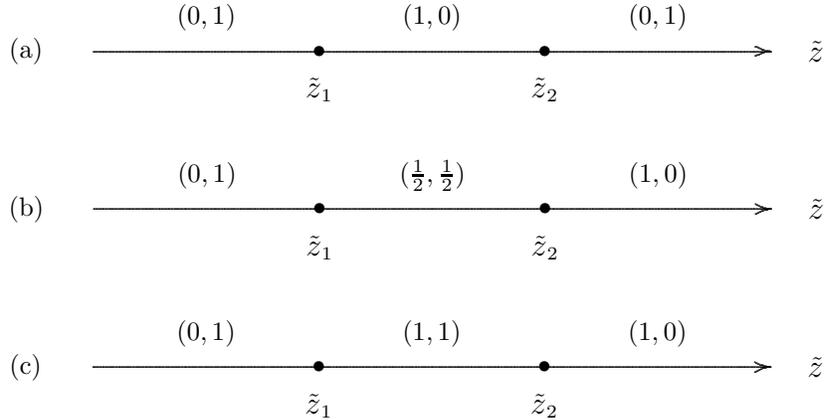}
\caption{The rod structure of: (a) the Euclidean Schwarzschild and Kerr instantons; (b) the Eguchi--Hanson and double-centered Taub-NUT instantons; and (c) the Taub-bolt instanton; all in standard orientation.}
\end{center}
\end{figure}

\newsubsection{Euclidean Kerr instanton}

The Euclidean Kerr instanton \cite{Gibbons:1976} has the metric
\be
\label{Euclidean Kerr}
{\dif s}^{2}={\frac {{\Delta} \left( {\dif\psi}+a \sin ^{2}
 \theta\, {\dif\phi} \right) ^{2}}{{\Sigma}}}+{\frac { \sin ^{2} \theta\,
 [a{\dif\psi}- \left( {r}^{2}-{a}^{2} \right) {\dif\phi}] ^{2}}{{\Sigma}}}+{\Sigma} \left( {\frac {{\dif r}
^{2}}{{\Delta}}}+{{\dif\theta}}^{2} \right),
\ee
where $\Sigma$ and $\Delta$ are defined as
\be
{\Sigma}={r}^{2}-{a}^{2} \cos ^{2} \theta \,,\qquad {\Delta}={r}^{2}-2mr-{a}^{2}\,.
\ee
The parameters $m$, $a$ and coordinates $r$, $\theta$ take the ranges $m\geq 0$, $-\infty <a<\infty$, $r\geq r_0$, $0\leq \theta \leq \pi$, where $r_0$ is defined as the larger root of $\Delta$, i.e., $r_0=m+\sqrt {{m}^{2}+{a}^{2}}$.

The Weyl--Papapetrou coordinates $(\psi,\phi,\rho,z)$ are related to the above coordinates by
\be
\label{Weyl_Kerr}
\rho=\sqrt {{r}^{2}-2mr-{a}^{2}}\,\sin \theta \,,\qquad z= \left( r-m \right) \cos \theta \,.
\ee
In these coordinates, the rod structure is similar to that of the Euclidean Schwarzschild instanton. There are two turning points, located at $(\rho=0, z=z_1\equiv-\sqrt {{m}^{2}+{a}^{2}})$ or $(r=r_0, \theta=\pi)$, and $(\rho=0, z=z_2\equiv\sqrt {{m}^{2}+{a}^{2}})$ or $(r=r_0, \theta=0)$. They divide the $z$-axis into three rods; from left to right they are:

\begin{itemize}
\item Rod 1: a semi-infinite rod located at $(\rho=0, z\leq z_1)$ or $(r\geq r_0, \theta=\pi)$, with direction $\ell_{1}=(0,1)$.

\item Rod 2: a finite rod located at $(\rho=0, z_1\leq z\leq z_2)$ or $(r=r_0, 0\leq\theta\leq\pi)$, with direction $\ell_{2}=(\frac{1}{\kappa_E}, \frac{\Omega_E}{\kappa_E})$, where $\kappa_E$ is the Euclidean surface gravity on the horizon and $\Omega_E$ is the Euclidean angular velocity of the horizon, given by
\be
\kappa_E={\frac {\sqrt {{m}^{2}+{a}^{2}}}{2m (m+
\sqrt {{m}^{2}+{a}^{2}})}}\,,\qquad
\Omega_E={\frac {a}{2m (m+\sqrt {{m}^{2}+{a}^{2}})}}\,.
\ee

\item Rod 3: a semi-infinite rod located at $(\rho=0, z\geq z_2)$ or $(r\geq r_0, \theta=0)$, with direction $\ell_{3}=(0,1)$.
\end{itemize}

It is straightforward to check that the direction pairs $\{\ell_1,\ell_2\}$ and $\{\ell_2,\ell_3\}$ of adjacent rods are related by a $GL(2,\mathbb{Z})$ transformation. To ensure regularity, the orbits generated by say the first direction pair $\{\ell_1,\ell_2\}$ should be identified with period $2\pi$ independently, i.e.,
\be
\label{Identifications for EK}
(\psi, \phi) \rightarrow (\psi, \phi+2 \pi) \,,\qquad (\psi, \phi) \rightarrow \left(\psi+\frac{2\pi}{\kappa_E}, \phi+ \frac{2\pi\Omega_E}{\kappa_E}\right) \,.
\ee
The direction pair $\{\ell_1,\ell_2\}$, or equivalently $\{\ell_2,\ell_3\}$, is then identified as the pair of independent $2\pi$-periodic generators of the $U(1)\times U(1)$ isometry group of the Euclidean Kerr instanton.

Obviously, when $a\rightarrow 0$ we recover the Euclidean Schwarzschild instanton from the Euclidean Kerr instanton. On the other hand, when $m\rightarrow 0$ we recover a three-dimensional flat space in a new form; this is similar to the $m\rightarrow 0$ limit of the Euclidean Schwarzschild instanton.

If $m\neq 0$, we can put the rod structure in standard orientation by taking $\{\tilde{V}_{(1)}=\ell_2,\tilde{V}_{(2)}=\ell_1\}$. The corresponding new Weyl--Papapetrou coordinates $(\tilde{\psi},\tilde{\phi},\tilde{\rho},\tilde{z})$ are related to the old coordinates (\ref{Weyl_Kerr}) by
\be
\psi=\frac{1}{\kappa_E}\, \tilde{\psi} \,,\qquad
\phi=\frac{\Omega_E}{\kappa_E}\, \tilde{\psi}+\tilde{\phi} \,,\qquad
\rho=\kappa_E \tilde{\rho}\,,\qquad
z=\kappa_E \tilde{z}\,.
\ee
The two turning points are now pushed to $(\tilde{\rho}=0,\tilde{z}=\tilde{z}_1\equiv-2 m (m+\sqrt{m^2+a^2}))$ and $(\tilde{\rho}=0,\tilde{z}=\tilde{z}_2\equiv2 m (m+\sqrt{m^2+a^2}))$, and the corresponding directions of the three rods from left to right are $K_1=(0,1)$, $K_2=(1,0)$ and $K_3=(0,1)$. This is illustrated in Fig.~2(a). In the new Weyl--Papapetrou coordinates, the following identifications should be made to ensure regularity:
\be
\label{Identifications for EK 2}
(\tilde{\psi}, \tilde{\phi}) \rightarrow (\tilde{\psi},\tilde{ \phi}+2 \pi) \,,\qquad (\tilde{\psi}, \tilde{\phi}) \rightarrow (\tilde{\psi}+2 \pi, \tilde{\phi}) \,.
\ee

As in the case of the Euclidean Schwarzschild instanton, the topology of the constant $r>r_0$ surfaces is $S^1\times S^2$, with $\ell_2$ and $\ell_1$ being the generators of the rotations for $S^1$ and $S^2$ respectively. When $r\rightarrow r_0$, the $S^1$ vanishes, and the constant $r$ surface becomes a two-sphere $S^2$ at $r=r_0$. But unlike the situation in the Euclidean Schwarzschild instanton, the $S^1$ blows up at infinity $r\rightarrow \infty$ in the general case for $a\neq 0$ \cite{Aharony:2002}, which can be seen from the fact that $g_{\tilde{\psi} \tilde{\psi}}$ diverges at infinity. However, in this general case, if $-1<\frac{\Omega_E}{\kappa_E}<1$ is a rational number $\frac{q}{p}$ with coprime integers $p$ and $q$, then the Killing vector field $\frac{\partial}{\partial{\psi}}$ generates closed and finite orbits with period $\frac{2\pi p}{\kappa_E}$ at infinity \cite{Hunter:1998}. It can be checked that, like the Euclidean Schwarzschild instanton, the Euclidean Kerr instanton is AF.

The rod structure of the Euclidean Kerr instanton admits a one-parameter degeneracy, which can be seen from the fact that we can appropriately vary $m$ and $a$ such that the turning points $(\tilde{\rho}=0,\tilde{z}=\tilde{z}_1)$ and $(\tilde{\rho}=0,\tilde{z}=\tilde{z}_2)$ remain fixed. We also note that there is a one-parameter family of the Euclidean Kerr instanton that has exactly the same rod structure (in standard orientation) as the Euclidean Schwarzschild instanton. This is another example of the fact that the rod structure alone cannot uniquely determine a solution, even when NUT charge is absent. And again this does not violate Hollands and Yazadjiev's theorem \cite{Hollands:2008}, as the Euclidean Kerr instanton has an $S^1$ circle whose radius diverges at infinity, and moreover its metric has a cross term $g_{\tilde{\psi}\tilde{\phi}}$ that diverges at infinity.

\newsubsection{Eguchi--Hanson instanton}

The Eguchi--Hanson instanton \cite{Eguchi:1978} has the well-known metric
\be
\label{Eguchi-Hanson}
{\dif s}^{2}=\left( 1-{\frac {{a}^{4}}{{r}^{4}}} \right) \frac{{r}^{
2}}{4} \left( {\dif\psi}+\cos \theta\, {\dif\phi} \right) ^{2
}+\left( 1-{\frac {{a}^{4}}{{r}^{4}}} \right) ^{-1} {\dif r}^{2}+\frac{{r}^{2}}{4} ( {{\dif\theta}}^{2}+ \sin^{2} \theta\,
 {{\dif\phi}}^{2} )\,,
\ee
where the parameter $a$ and coordinates $r$, $\theta$ take the ranges $r\geq a>0$, $0\leq \theta \leq \pi$.

The Weyl--Papapetrou coordinates $(\psi,\phi,\rho,z)$ are related to the above coordinates by
\be
\label{Weyl_EH}
\rho=\frac{1}{4}\,\sqrt {{r}^{4}-{a}^{4}}\,\sin \theta \,,\qquad z=\frac{{r}^{2}}{4}\cos \theta \,.
\ee
In these coordinates, the rod structure has two turning points at $(\rho=0, z=z_1\equiv-\frac{a^2}{4})$ or $(r=a, \theta=\pi)$, and at $(\rho=0, z=z_2\equiv\frac{a^2}{4})$ or $(r=a, \theta=0)$. It consists of the following three rods:

\begin{itemize}
\item Rod 1: a semi-infinite rod located at $(\rho=0, z\leq z_1)$ or $(r\geq a, \theta=\pi)$, with direction $\ell_{1}=(1,1)$.

\item Rod 2: a finite rod located at $(\rho=0, z_1\leq z\leq z_2)$ or $(r=a, 0\leq\theta\leq\pi)$, with direction $\ell_{2}=(1,0)$.

\item Rod 3: a semi-infinite rod located at $(\rho=0, z\geq z_2)$ or $(r\geq a, \theta=0)$, with direction $\ell_{3}=(-1,1)$.
\end{itemize}

It is straightforward to check that the direction pairs $\{\ell_1,\ell_2\}$ and $\{\ell_2,\ell_3\}$ of adjacent rods are related by a $GL(2,\mathbb{Z})$ transformation. To ensure regularity, the orbits generated by say the first direction pair $\{\ell_1,\ell_2\}$ should be identified with period $2\pi$ independently. This direction pair can then be taken as the pair of independent $2\pi$-periodic generators of the $U(1)\times U(1)$ isometry group of the Eguchi--Hanson instanton. However, since the pair $\{\ell_1-\ell_2,\ell_2\}$ is also related to $\{\ell_1,\ell_2\}$ by a $GL(2,\mathbb{Z})$ transformation, and since $\ell_1-\ell_2=(0,1)$, we can make the following identifications to ensure regularity:
\be
(\psi, \phi) \rightarrow (\psi, \phi+2 \pi) \,,\qquad (\psi, \phi) \rightarrow (\psi+2 \pi, \phi) \,.
\ee
Note that in this case, the direction pair $\{\ell_1,\ell_3\}$ corresponding to the two semi-infinite rods is not related to $\{\ell_1,\ell_2\}$ by a $GL(2,\mathbb{Z})$ transformation, so it cannot be taken as the pair of independent $2\pi$-periodic generators of the $U(1)\times U(1)$ isometry group.

We can put the rod structure in standard orientation by taking $\{\tilde{V}_{(1)}=-\ell_3,\tilde{V}_{(2)}=\ell_1\}$. The corresponding new Weyl--Papapetrou coordinates $(\tilde{\psi},\tilde{\phi},\tilde{\rho},\tilde{z})$ are related to the old coordinates (\ref{Weyl_EH}) by
\be
\label{new Weyl EH}
\psi=\tilde{\psi}+\tilde{\phi} \,,\qquad
\phi=-\tilde{\psi}+\tilde{\phi}\,,\qquad
\rho=\frac{1}{2}\tilde{\rho}\,,\qquad
z=\frac{1}{2}\tilde{z}\,.
\ee
The two turning points are now pushed to $(\tilde{\rho}=0,\tilde{z}=\tilde{z}_1\equiv-\frac{a^2}{2})$ and $(\tilde{\rho}=0,\tilde{z}=\tilde{z}_2\equiv\frac{a^2}{2})$, and the corresponding directions of the three rods from left to right are $K_1=(0,1)$, $K_2=(\frac{1}{2},\frac{1}{2})$ and $K_3=(1,0)$.\footnote{Strictly speaking, we should have $K_3=(-1,0)$ to keep the direction of rod 3 invariant under the coordinate transformation (\ref{new Weyl EH}). But recall that we allow a possible minus sign for the direction of a rod, so here we take $K_3=(1,0)$.} This is illustrated in Fig.~2(b). In the new Weyl--Papapetrou coordinates, the following identifications should be made to ensure regularity:
\be
\label{identification for Eguchi-Hanson}
(\tilde{\psi}, \tilde{\phi}) \rightarrow (\tilde{\psi}, \tilde{\phi}+2 \pi) \,,\qquad (\tilde{\psi}, \tilde{\phi}) \rightarrow (\tilde{\psi}+\pi,\tilde{ \phi}+\pi) \,.
\ee

Surfaces of constant $r>a$ are represented in the $(\rho,z)$ half-plane of the orbit space by curves intersecting with rods 1 and 3. Since $K_3=2 K_2-K_1$, the topology of these surfaces is then a lens-space $L(2,-1)\cong\mathbb{R}P^3$. They shrink down to a two-sphere at $r=a$. At infinity $r\rightarrow \infty$, the Eguchi--Hanson instanton approaches a four-dimensional flat space quotiented by a $\mathbb{Z}_2$ group, i.e., $E^4/\mathbb{Z}_2$. It can be checked that this instanton is ALE with $\Gamma=\mathbb{Z}_2$.

Finally, we note that, under the above identifications, the limit $a\rightarrow 0$ results in a four-dimensional flat space quotiented by a $\mathbb{Z}_2$ group, which is singular as there is an orbifold singularity present at the origin.

\newsubsection{Double-centered Taub-NUT instanton}

The double-centered Taub-NUT instanton \cite{Hawking:1976,Gibbons:1979} has the metric
\be
\label{double-centered Gibbons-Hawking}
{\dif s}^{2}=H^{-1} { \left( {\dif\psi}+A \right) ^{2}}+H
 ( {\dif r}^{2}+{r}^{2} {{\dif\theta}}^{2}+{r}^{2} \sin^{2}
 \theta\, {{\dif\phi}}^{2} )\,,
\ee
where the harmonic function $H$ and the twist potential $A$ on $E^3$ are defined as
\ba
H&=&1+{\frac {2 |n_1|}{{r_1}}}+{\frac {2 |n_2|}{{r_2}}}\,,\cr
A&=&{\frac {2 n_1 \left( r\cos \theta  +a \right) }{{r_1}}}\,{\dif\phi
}+{\frac {2 n_2 \left( r\cos \theta -a
 \right) }{{r_2}}}\,{\dif\phi}\,,
\ea
with $r_1=\sqrt {{r}^{2}+{a}^{2}+2ar\cos  \theta  }$ and $r_2=\sqrt {{r}^{2}+{a}^{2}-2ar\cos  \theta  }$. The parameters $a$, $n_1$, $n_2$ and coordinates $r$, $\theta$ take the ranges $a>0$, $-\infty <n_1,n_2<\infty$, $r\geq 0$, $0\leq \theta \leq \pi$. Furthermore, the NUT charges $n_1$ and $n_2$ are required to have the same sign.

The Weyl--Papapetrou coordinates $(\psi,\phi,\rho,z)$ are related to the above coordinates by
\be
\label{Weyl_double_TN}
\rho=r\sin \theta  \,,\qquad z=r\cos \theta \,.
\ee
In these coordinates, the rod structure has two turning points at $(\rho=0, z=z_1\equiv-a)$ or $(r=a, \theta=\pi)$, and at $(\rho=0, z=z_2\equiv a)$ or $(r=a, \theta=0)$. It consists of the following three rods:

\begin{itemize}
\item Rod 1: a semi-infinite rod located at $(\rho=0, z\leq z_1)$ or $(r\geq a, \theta=\pi)$, with direction $\ell_{1}=(2n_1+2n_2,1)$.

\item Rod 2: a finite rod located at $(\rho=0, z_1\leq z\leq z_2)$ or $(0 \leq r \leq a,\, \theta=0 \hspace{1.5mm} {\rm and} \hspace{1.5mm}  \pi)$, with direction $\ell_{2}=(-2n_1+2n_2,1)$.

\item Rod 3: a semi-infinite rod located at $(\rho=0, z\geq z_2)$ or $(r\geq a, \theta=0)$, with direction $\ell_{3}=(-2n_1-2n_2,1)$.
\end{itemize}

It can be checked that the direction pairs $\{\ell_1,\ell_2\}$ and $\{\ell_2,\ell_3\}$ of adjacent rods are related by a $GL(2,\mathbb{Z})$ transformation, provided the two NUT charges are equal: $n_1=n_2\equiv n$; this is what we assume from now on. To further ensure regularity, the orbits generated by say the first direction pair $\{\ell_1,\ell_2\}$ should be identified with period $2\pi$ independently. This direction pair can then be taken as the pair of independent $2\pi$-periodic generators of the $U(1)\times U(1)$ isometry group of the double-centered Taub-NUT instanton. However, since the pair $\{\ell_1-\ell_2,\ell_2\}$ is also related to $\{\ell_1,\ell_2\}$ by a $GL(2,\mathbb{Z})$ transformation, and since $\ell_1-\ell_2=(4n,0)$, we can make the following identifications to ensure regularity:
\be
\label{identification for double TN}
(\psi, \phi) \rightarrow (\psi+8n \pi, \phi) \,,\qquad (\psi, \phi) \rightarrow (\psi, \phi+2 \pi) \,.
\ee
As in the Eguchi--Hanson instanton, the direction pair $\{\ell_1,\ell_3\}$ corresponding to the two semi-infinite rods is not related to $\{\ell_1,\ell_2\}$ by a $GL(2,\mathbb{Z})$ transformation, so it cannot be taken as the pair of independent $2\pi$-periodic generators of the $U(1)\times U(1)$ isometry group.

If $n\neq 0$, we can put the rod structure in standard orientation by taking $\{\tilde{V}_{(1)}=\ell_3,\tilde{V}_{(2)}=\ell_1\}$. The corresponding new Weyl--Papapetrou coordinates $(\tilde{\psi},\tilde{\phi},\tilde{\rho},\tilde{z})$ are related to the old coordinates (\ref{Weyl_double_TN}) by
\be
\psi=-4 n (\tilde{\psi}-\tilde{\phi}) \,,\qquad
\phi=\tilde{\psi}+\tilde{\phi} \,,\qquad
\rho=\frac{1}{8|n|}\tilde{\rho}\,,\qquad
z=\frac{1}{8|n|}\tilde{z}\,.
\ee
The two turning points are now pushed to $(\tilde{\rho}=0,\tilde{z}=\tilde{z}_1\equiv-8|n|a)$ and $(\tilde{\rho}=0,\tilde{z}=\tilde{z}_2\equiv8|n|a)$, and the corresponding directions of the three rods from left to right are $K_1=(0,1)$, $K_2=(\frac{1}{2},\frac{1}{2})$ and $K_3=(1,0)$. This is illustrated in Fig.~2(b). In the new Weyl--Papapetrou coordinates, the following identifications should be made to ensure regularity:
\be
(\tilde{\psi}, \tilde{\phi}) \rightarrow (\tilde{\psi}, \tilde{\phi}+2 \pi) \,,\qquad (\tilde{\psi}, \tilde{\phi}) \rightarrow (\tilde{\psi}+\pi,\tilde{ \phi}+\pi) \,.
\ee

It is easy to show that constant $r$ surfaces have the topology $S^1\times S^2$ for $0<r<a$, and the topology $L(2,1)$ for $r>a$. At infinity $r\rightarrow \infty$, the Killing vector field $\frac{\partial}{\partial \psi}$ generates a compact dimension of constant size $8n\pi$. It can be checked that the double-centered Taub-NUT instanton is ALF with $\Gamma=\mathbb{Z}_2$. When $n\rightarrow0$, we recover a three-dimensional flat space; on the other hand, when $n\rightarrow\infty$, the finite dimension blows up, and we recover the Eguchi--Hanson instanton.

As can be seen, the above rod structure also admits a one-parameter degeneracy. We also note that there is a one-parameter family of the double-centered Taub-NUT instanton (\ref{double-centered Gibbons-Hawking}) that has exactly the same rod structure (in standard orientation) as the Eguchi--Hanson instanton (\ref{Eguchi-Hanson}). This is the third example that the rod structure alone cannot uniquely determine a solution. Actually the double-centered Taub-NUT instanton generalises the Eguchi--Hanson instanton, in the same way that the self-dual Taub-NUT instanton generalises four-dimensional flat space. So the same kind of degeneracy of the rod structure appears.

\newsubsection{Taub-bolt instanton}

The Euclidean non-self-dual Taub-NUT solution has the metric
\be
\label{non-self-dual Taub-NUT}
{\dif s}^{2}=f(r)  \left( {\dif\psi}+2n\cos
 \theta\, {\dif\phi} \right) ^{2}+{\frac {{\dif r}^{2
}}{f(r) }}+ \left( {r}^{2}-{n}^{2} \right)  ( {{
\dif\theta}}^{2}+\sin ^{2} \theta \,{{
\dif\phi}}^{2} )\,,
\ee
where the function $f(r)$ is defined as
\be
f(r) ={\frac {{r}^{2}+{n}^{2}-2mr}{{r}^{2}-{n}^{2}}}\,.
\ee
The parameters $m$, $n$ and coordinates $r$, $\theta$ take the ranges $m\geq|n|$, $r\geq r_0$, $0\leq \theta \leq \pi$, where $r_0$ is the larger root of $f(r)$, i.e., $r_0=m+\sqrt{m^2-n^2}$. The self-dual Taub-NUT metric (\ref{self-dual Taub-NUT}) can be recovered from (\ref{non-self-dual Taub-NUT}) by setting $m=|n|$.

The Weyl--Papapetrou coordinates $(\psi,\phi,\rho,z)$ are related to the above coordinates by
\be
\label{Weyl_TB}
\rho=\sqrt {{r}^{2}-2m r+{n}^{2}}\,\sin \theta \,,\qquad z= \left( r-m \right) \cos \theta \,.
\ee
In these coordinates, the rod structure has two turning points at $(\rho=0, z=z_1\equiv-\sqrt{m^2-n^2})$ or $(r=r_0, \theta=\pi)$, and at $(\rho=0, z=z_2\equiv\sqrt{m^2-n^2})$ or $(r=r_0, \theta=0)$. It consists of the following three rods:

\begin{itemize}
\item Rod 1: a semi-infinite rod located at $(\rho=0, z\leq z_1)$ or $(r\geq r_0, \theta=\pi)$, with direction $\ell_{1}=(2 n,1)$.

\item Rod 2: a finite rod located at $(\rho=0, z_1\leq z\leq z_2)$ or $(r=r_0, 0\leq\theta\leq\pi)$, with direction $\ell_{2}=(2m+2\sqrt{m^2-n^2},0)$.

\item Rod 3: a semi-infinite rod located at $(\rho=0, z\geq z_2)$ or $(r\geq r_0, \theta=0)$, with direction $\ell_{3}=(-2 n,1)$.
\end{itemize}

It can be checked that the direction pairs $\{\ell_1,\ell_2\}$ and $\{\ell_2,\ell_3\}$ of adjacent rods are related by a $GL(2,\mathbb{Z})$ transformation, provided $m=\frac{5}{4}\,|n|$. This corresponds to the Taub-bolt instanton discovered by Page \cite{Page:1979}. In what follows, we focus on the Taub-bolt instanton only; in this case, we can take $\ell_{2}=(4n,0)$ without loss of generality. To further ensure regularity, the orbits generated by say the first direction pair $\{\ell_1,\ell_2\}$ should be identified with period $2\pi$ independently, i.e.,
\be
\label{identification for TB}
(\psi, \phi) \rightarrow (\psi+4 n \pi, \phi+2 \pi)\,,\qquad (\psi, \phi) \rightarrow (\psi+8 n \pi, \phi) \,.
\ee
These identifications are, in fact, the same as (\ref{identification for self-dual Taub-NUT 2}) needed to make the self-dual Taub-NUT instanton (\ref{self-dual Taub-NUT}) regular. This is because the direction pair $\{\ell_1,\ell_3\}$ corresponding to the two semi-infinite rods is related to the direction pair $\{\ell_1,\ell_2\}$ by a $GL(2,\mathbb{Z})$ transformation, and so can be taken as the pair of independent $2\pi$-periodic generators of the $U(1)\times U(1)$ isometry group of the Taub-bolt instanton.

If $n\neq 0$, we can put the rod structure in standard orientation by taking $\{\tilde{V}_{(1)}=-\ell_3,\tilde{V}_{(2)}=\ell_1\}$. The corresponding new Weyl--Papapetrou coordinates $(\tilde{\psi},\tilde{\phi},\tilde{\rho},\tilde{z})$ are related to the old coordinates (\ref{Weyl_TB}) by
\be
\psi=2n (\tilde{\psi}+\tilde{\phi}) \,,\qquad
\phi=-\tilde{\psi}+\tilde{\phi} \,,\qquad
\rho=\frac{1}{4|n|}\tilde{\rho}\,,\qquad
z=\frac{1}{4|n|}\tilde{z}\,.
\ee
The two turning points are now pushed to $(\tilde{\rho}=0,\tilde{z}=\tilde{z}_1\equiv-3n^2)$ and $(\tilde{\rho}=0,\tilde{z}=\tilde{z}_2\equiv3n^2)$, and the corresponding directions of the three rods from left to right are $K_1=(0,1)$, $K_2=(1,1)$ and $K_3=(1,0)$.\footnote{As in the case of the Eguchi--Hanson instanton, we should strictly speaking have $K_3=(-1,0)$, but here we take $K_3=(1,0)$ without making any difference.} This is illustrated in Fig.~2(c). Since $\{K_1,K_3\}$ can be taken as the pair of independent $2\pi$-periodic generators of the $U(1)\times U(1)$ isometry group, in the new Weyl--Papapetrou coordinates the following identifications can be made to ensure regularity:
\be
\label{identification for TB 2}
(\tilde{\psi}, \tilde{\phi}) \rightarrow (\tilde{\psi}, \tilde{\phi}+2 \pi) \,,\qquad (\tilde{\psi}, \tilde{\phi}) \rightarrow (\tilde{\psi}+2\pi,\tilde{ \phi}) \,.
\ee
Note that the Eguchi--Hanson and double-centered Taub-NUT instantons share the same Harmark rod structure \cite{Harmark:2004,Harmark:2005} as the Taub-bolt instanton (in appropriate coordinates), but the stronger form of the rod structure as defined in this paper is able to distinguish the two cases.

It is easy to show that surfaces of constant $r>r_0$ have topology $S^3$, and that they shrink down to a two-sphere $S^2$ at $r=r_0$. At infinity $r\rightarrow \infty$, the Taub-bolt instanton approaches a finite $S^1$ fibre bundle over an $S^2$, which is the Hopf fibration of $S^3$. The Killing vector field $\frac{\partial}{\partial \psi}$ generates the finite $S^1$ fibre at infinity, with a constant size $8n\pi$. Like the self-dual Taub-NUT instanton, the Taub-bolt instanton is ALF with $\Gamma=1$. When $n\rightarrow0$, we recover a three-dimensional flat space.

Finally, we point out that the Eguchi--Hanson metric (\ref{Eguchi-Hanson}) can be recovered from the non-self-dual Taub-NUT metric (\ref{non-self-dual Taub-NUT}) up to coordinate transformations, by taking the limit $m\rightarrow \infty$ with $m^4-n^4$ fixed.

\newsubsection{No completely regular Kerr-bolt instanton}

The Kerr-bolt instanton was first discussed by Gibbons and Perry \cite{Gibbons:1979b}. It generalises the Taub-bolt instanton in the same way that the Kerr solution generalises the Schwarzschild solution, and was obtained as a special case of the Euclidean Kerr-NUT metric. Here we take a simpler form of the Euclidean Kerr-NUT metric as used in Ghezelbash et al.~\cite{Ghezelbash:2007}:
\ba
\label{Euclidean Kerr-bolt}
{\dif s}^{2}&=&{\frac {{\Delta}}{{\Sigma}}} [{\dif\psi}+ \left( 2n
\cos \theta +a \sin  ^{2} \theta
  \right) {\dif\phi}] ^{2}+{\frac {
  \sin ^{2} \theta  }{{\Sigma}}}[a{\dif\psi}-
 \left( {r}^{2}-{n}^{2}-{a}^{2} \right) {\dif\phi}] ^{2}\cr
 &&+{\Sigma} \left( {\frac {{\dif r}^{2}}{{\Delta}}}
+{{\dif\theta}}^{2} \right),
\ea
where $\Sigma$ and $\Delta$ are defined as
\be
{\Sigma}={r}^{2}- \left( n-a\cos \theta  \right) ^{
2},\qquad
{\Delta}={r}^{2}-2mr-{a}^{2}+{n}^{2}.
\ee
The parameters $m$, $n$, $a$ and coordinates $r$, $\theta$ take the ranges $m\geq|n|$, $-\infty<a<\infty$, $r\geq r_0$, $0\leq \theta \leq \pi$, where $r_0$ is the larger root of $\Delta$, i.e., $r_0=m+\sqrt {{m}^{2}+{a}^{2}-{n}^{2}}$. The Euclidean Kerr metric (\ref{Euclidean Kerr}) and the non-self-dual Taub-NUT metric (\ref{non-self-dual Taub-NUT}) can be recovered from (\ref{Euclidean Kerr-bolt}) by setting $n=0$ and $a=0$, respectively.

The Weyl--Papapetrou coordinates $(\psi,\phi,\rho,z)$ are related to the above coordinates by
\be
\rho=\sqrt {{r}^{2}-2mr-{a}^{2}+{n}^{2}}\,\sin \theta \,,\qquad
z= \left( r-m \right) \cos \theta \,.
\ee
In these coordinates, the rod structure has two turning points, at $(\rho=0, z=z_1\equiv-\sqrt {{m}^{2}+{a}^{2}-{n}^{2}})$ or $(r=r_0, \theta=\pi)$, and at $(\rho=0, z=z_2\equiv\sqrt {{m}^{2}+{a}^{2}-{n}^{2}})$ or $(r=r_0, \theta=0)$. It consists of the following three rods:

\begin{itemize}
\item Rod 1: a semi-infinite rod located at $(\rho=0, z\leq z_1)$ or $(r\geq r_0, \theta=\pi)$, with direction $\ell_{1}=(2 n,1)$.

\item Rod 2: a finite rod located at $(\rho=0, z_1\leq z\leq z_2)$ or $(r=r_0, 0\leq\theta\leq\pi)$, with direction $\ell_{2}=(\frac{1}{\kappa_E},\frac{\Omega_E}{\kappa_E})$, where $\kappa_E$ and $\Omega_E$ are defined as
\be
\kappa_E={\frac {\sqrt {{m}^{2}+{a}^{2}-{n}^{2}}}{2({m}^{2}-{n}^{2}+m
\sqrt {{m}^{2}+{a}^{2}-{n}^{2}})}}\,,\qquad
\Omega_E={\frac {a}{2({m}^{2}-{n}^{2}+m\sqrt {{m}^{2}+{a}^{2}-{n}^{2}
})}}\,.
\ee

\item Rod 3: a semi-infinite rod located at $(\rho=0, z\geq z_2)$ or $(r\geq r_0, \theta=0)$, with direction $\ell_{3}=(-2 n,1)$.
\end{itemize}

Without loss of generality, we consider the case $0\leq n \leq m$; a similar analysis applies when $-m\leq n \leq 0$. When $n=0$, we recover the Euclidean Kerr instanton, in which case $\ell_1=\ell_3$. When $n=m$, we have $\ell_1=\ell_2$ and thus in this case the first and the second rods actually join up to form a single rod, leaving only one turning point in the rod structure. The resulting metric is nothing but the self-dual Taub-NUT metric (\ref{self-dual Taub-NUT}) in a different form.

When $0< n < m$, all three directions are mutually linearly independent. If we impose the condition that the direction pairs $\{\ell_1,\ell_2\}$ and $\{\ell_2,\ell_3\}$, of adjacent rods intersecting at the two turning points, are related by a $GL(2,\mathbb{Z})$ transformation, it is straightforward to show that the only solution is $\{a=0,m=\frac{5n}{4}\}$, which is just the Taub-bolt instanton. This result implies that no new gravitational instantons can be obtained from (\ref{Euclidean Kerr-bolt}) when $a\neq0$; in particular, there is no Kerr-bolt instanton satisfying our regularity conditions!

It is worth examining this result in more detail, especially in the light of previous analyses of the Kerr-bolt instanton. In order to avoid conical singularities along all three rods, the orbits generated by $\ell_1$, $\ell_2$ and $\ell_3$ should be respectively identified with period $2\pi$; at the same time, it is necessary for the so-called compatibility condition $q\ell_1+p\ell_2+s\ell_3=0$ to be satisfied for mutually coprime integers $p$, $q$ and $s$ (see e.g., \cite{Ghezelbash:2007} for an explanation of this and some information on lattice analysis). This class of Kerr-bolt instantons free of conical singularities was analyzed in detail in \cite{Ghezelbash:2007}. However, it turns out that they contain orbifold singularities in general. We recall that, by first identifying the orbits generated by $\ell_1$ and $\ell_2$ with period $2\pi$ respectively, the metric in the vicinity of the first turning point $(\rho=0,z=z_1)$ can be brought into the standard form of $E^4$ near the origin (\ref{R4 origin}), with $\ell_1=\frac{\partial}{\partial \phi_1}$ and $\ell_2=\frac{\partial}{\partial \phi_2}$. Since $\ell_3=-\frac{1}{s}{(q\ell_1+p\ell_2)}$, if we further identify the orbits generated by $\ell_3$ with period $2\pi$, we quotient the space by a $\mathbb{Z}_{|s|}$ identification group. This will leave a $\mathbb{Z}_{|s|}$ orbifold singularity at the first turning point if $|s|\geq 2$. Similarly, if $|q|\geq 2$ there will be a $\mathbb{Z}_{|q|}$ orbifold singularity at the second turning point $(\rho=0,z=z_2)$.\footnote{Moreover, the presence of these orbifold singularities would imply that the bolt at $r=r_0$ itself will not be a completely regular $S^2$ surface. This surface will in general possess conical singularities at the two poles.} If we require the absence of these two orbifold singularities, we have to impose $s=\pm 1$ and $q=\pm 1$. It is easy to see that the only solution to these conditions is $\{a=0,m=\frac{5n}{4}\}$, i.e., the Taub-bolt instanton.

Hence, there does not exist a completely regular Kerr-bolt instanton free of both conical and orbifold singularities. The space with metric (\ref{Euclidean Kerr-bolt}) is regular only in two special cases, the Euclidean Kerr instanton and the Taub-bolt instanton, as described above. However, we should also note that the local metrics of the self-dual Taub-NUT and Eguchi--Hanson instantons can also be recovered from (\ref{Euclidean Kerr-bolt}) up to coordinate transformations, by taking very special limits.

\newsubsection{Multi-collinearly-centered Taub-NUT instanton}

The general multi-Taub-NUT instanton \cite{Hawking:1976,Gibbons:1979} admits only a one-parameter isometry group, but a particular class admits a larger $U(1)\times U(1)$ isometry group, namely those with all the nuts collinearly centered on the base space $E^3$. The metric of this special class takes the form
\be
\label{multi Gibbons-Hawking}
{\dif s}^{2}= H^{-1} \left( {\dif\psi}+A \right) ^{2}+H
 ( {\dif r}^{2}+{r}^{2} {{\dif\theta}}^{2}+ r^2 \sin ^{2}
 \theta\, {{\dif\phi}}^{2}  )\,,
\ee
where the harmonic function $H$ and the twist potential $A$ on $E^3$ are defined as
\ba
H&=&1+\sum_{i=1}^k \frac{2|n_i|}{r_i}\,,\cr
A&=& \sum_{i=1}^k \frac{2 n_i (r \cos \theta-a_i)}{r_i}\,\dif\phi\,.
\ea
Here $r_i=\sqrt {{r}^{2}+{{a_i}}^{2}-2 {a_i} r \cos \theta}$ is the distance between the $i$-th nut and the position under consideration on the flat base space $E^3$ parameterised by the spherical polar coordinates $(r,\theta,\phi)$. Without loss of generality, we assume $0<a_1<\cdots<a_k$. There are in total $k$ nuts in the space, collinearly centered on the base space $E^3$ along the symmetry axis at $(r=a_i,\theta=0)$ respectively. The parameters $n_i$ and coordinates $r$, $\theta$ take the ranges $-\infty<n_i<\infty$, $r\geq 0$, $0\leq \theta \leq \pi$. Furthermore, all the NUT charges $n_i$ are required to have the same sign.

The Weyl--Papapetrou coordinates $(\psi,\phi,\rho,z)$ are related to the above coordinates by
\be
\label{Weyl_multi_TN}
\rho=r\sin \theta\,,\qquad z=r\cos \theta \,.
\ee
In these coordinates, the rod structure has $k$ turning points in total, corresponding to the $k$ nuts, located at $(\rho=0,z=z_i\equiv a_i)$ or $(r=a_i, \theta=0)$. The direction of the $i$-th rod is $\ell_i=(-2\displaystyle\sum_{j=1}^{i-1}{n_j}+2\sum_{j=i}^{k}{n_j},1)$, for $i=1,2,\dots,k+1$.

It can be checked that the condition for the adjacent rod direction pairs \{$\ell_{i-1}$,$\ell_i$\} and \{$\ell_i$,$\ell_{i+1}$\}, for $i=2,\dots,k$, to be related by a $GL(2,\mathbb{Z})$ transformation requires that $n_{i-1}=n_i\equiv n$. Hence, regularity of the gravitational instanton requires all the NUT charges to be equal \cite{Gibbons:1979c}; this is what we assume from now on. Furthermore, the orbits generated by say the $i$-th adjacent direction pair $\{\ell_i,\ell_{i+1}\}$ should be identified with period $2\pi$ independently. This direction pair is then identified as the pair of independent $2\pi$-periodic generators of the $U(1)\times U(1)$ isometry group of the multi-collinearly-centered Taub-NUT instanton. If $k$ is odd, we have $\ell_{\lfloor\frac{k}{2}\rfloor+1}=(2 n,1)$ and $\ell_{\lfloor\frac{k}{2}\rfloor+2}=(-2 n,1)$, and the identifications (\ref{identification for self-dual Taub-NUT 2}) should be made to ensure regularity; if $k$ is even, we have $\ell_{\frac{k}{2}}=(4 n,1)$ and $\ell_{\frac{k}{2}+1}=(0,1)$, and the identifications (\ref{identification for double TN}) should be made to ensure regularity.

If $n\neq 0$, we can put the rod structure in standard orientation by taking $\{\tilde{V}_{(1)}=\ell_{k+1},\tilde{V}_{(2)}=\ell_1\}$. The corresponding new Weyl--Papapetrou coordinates $(\tilde{\psi},\tilde{\phi},\tilde{\rho},\tilde{z})$ are related to the old coordinates (\ref{Weyl_multi_TN}) by
\be
\psi=-2 n k (\tilde{\psi}-\tilde{\phi}) \,,\qquad
\phi=\tilde{\psi}+\tilde{\phi}\,,\qquad
\rho=\frac{1}{4k|n|}\tilde{\rho}\,,\qquad
z=\frac{1}{4k|n|}\tilde{z}\,.
\ee
The $i$-th turning point is now pushed to $(\tilde{\rho}=0, \tilde{z}=\tilde{z}_i\equiv4|n|ka_i)$, and the directions of the $k+1$ rods from left to right are $K_1=(0,1)$, \dots, $K_i=(\frac{i-1}{k},\frac{k-i+1}{k})$, \dots, $K_{k+1}=(1,0)$. (Fig.~3 illustrates the case of $k=3$.) In the new Weyl--Papapetrou coordinates, the following identifications should be made to ensure regularity:
\be
(\tilde{\psi}, \tilde{\phi}) \rightarrow (\tilde{\psi}, \tilde{\phi}+2 \pi) \,,\qquad (\tilde{\psi}, \tilde{\phi}) \rightarrow \left(\tilde{\psi}+\frac{2\pi}{k},\tilde{ \phi}+\frac{2(k-1)\pi}{k}\right).
\ee
We note that the rod structure of the multi-collinearly-centered Taub-NUT instanton also admits a one-parameter degeneracy as that in the double-centered case discussed in Sec.~4.6.

\begin{figure}[t]
\begin{center}
\includegraphics{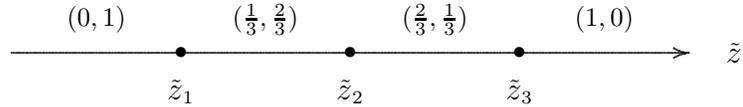}
\caption{The rod structure of the triple-collinearly-centered Taub-NUT instanton in standard orientation.}
\end{center}
\end{figure}

Constant $r$ surfaces for $a_i<r<a_{i+1}$ $(i=1,\dots,k-1)$ are represented in the $(\rho,z)$ half-plane of the orbit space by curves intersecting with the first and ${(i+1)}$-th rods, and since $K_{i+1}=-(i-1)K_1+i K_2$, these surfaces have lens-space topology $L(i,1-i)\cong L(i,1)$. By a similar argument, surfaces of constant $r>a_k$ have topology $L(k,1-k)\cong L(k,1)$. At infinity $r\rightarrow \infty$, the Killing vector field $\frac{\partial}{\partial \psi}$ generates a compact dimension of constant size $8n\pi$. It can be checked that the multi-collinearly-centered Taub-NUT instanton is ALF with $\Gamma=\mathbb{Z}_k$. When $n\rightarrow0$, we recover a three-dimensional flat space; on the other hand, when $n\rightarrow\infty$, the compact dimension blows up and we recover the collinearly-centered Gibbons--Hawking instanton \cite{Gibbons:1979}.\footnote{The usual form of the Gibbons--Hawking instanton is to omit the $1$ in the harmonic function $H$ \cite{Gibbons:1979}, but this is equivalent to the limit described above by taking $n\rightarrow \infty$.} The latter instanton is ALE with $\Gamma=\mathbb{Z}_k$.

The multi-collinearly-centered Taub-NUT instanton considered here, with all the axes at $\rho=0$ removed, is naturally taken as a $U(1)\times U(1)$-principal fibre bundle over the base space $\hat{I}=\{z+i\rho \in \mathbb{C}~|~ \rho>0 \}$. However, in the more conventional but equivalent way, with all the nuts removed, this gravitational instanton is taken as a $U(1)$-principal fibre bundle over the base space $E^3$ with the corresponding nut-points removed. See, e.g., \cite{Ida:2009} for this more conventional point of view.

\newsection{Possible new gravitational instantons}

\newsubsection{Possible new gravitational instantons with two turning points}
\label{sec:new gravitational instantons}

We have analyzed the rod structure of five classes of regular gravitational instantons with two turning points, namely the Euclidean Schwarzschild, Euclidean Kerr, Eguchi--Hanson, double-centered Taub-NUT and Taub-bolt instantons. It may be interesting to ask what kind of rod structure with two turning points is necessary for a regular solution.

As we have seen, any direction pair of adjacent rods, say $\{\ell_1,\ell_2\}$, should generate orbits with period $2\pi$ independently to avoid possible conical and orbifold singularities, and they are then identified as the pair of independent $2\pi$-periodic generators of the $U(1)\times U(1)$ isometry group of the space. Any other rod direction should be a linear combination of them with coprime integer coefficients, i.e., $\ell_3=q \ell_1+p \ell_2$ with coprime integers $p$ and $q$.\footnote{The case when $\ell_1=\ell_2$ or $\ell_2=\ell_3$ is trivial in our analysis, as it results in a rod structure with one turning point. We have already mentioned in Sec.~4.8 that the self-dual Taub-NUT metric can be recovered from the Kerr-bolt metric by imposing this condition.} Furthermore, the pair $\{\ell_2,\ell_3\}$ should also be related to the pair $\{\ell_1, \ell_2\}$ by a $GL(2,\mathbb{Z})$ transformation, so we have $q=\pm 1$. Depending on the value of $p$, we further divide the possibilities into two classes: (a) $p=0$, so $\ell_1=\pm \ell_3$, or (b) $|p|\geq 1$, so $\ell_2=(\ell_1\pm \ell_3)/p$. The first class corresponds to the case when the two semi-infinite rods are parallel, while the second class corresponds to the case when they are not.

\begin{figure}[t]
\begin{center}
\includegraphics{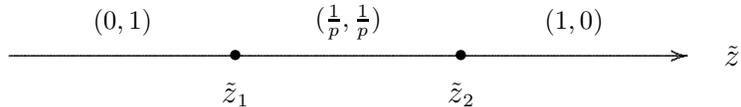}
\caption{The rod structure of possible new gravitational instantons with two turning points, for integer $|p|\geq3$, in standard orientation.}
\end{center}
\end{figure}

The Euclidean Schwarzschild and Kerr instantons are in class (a). The Taub-bolt instanton is in class (b) with $|p|=1$, while the Eguchi--Hanson and double-centered Taub-NUT instantons are in class (b) with $|p|=2$. If there exists a solution in class (b) with an integer $|p|\geq 3$ (the rod structure for such a solution in standard orientation is illustrated in Fig.~4), it would be possible to remove the conical and orbifold singularities, by identifying the orbits generated by say $\{\ell_1,\ell_2\}$ with period $2\pi$ independently. Asymptotically it will have a lens-space $L(p,1)$ structure, with or without any compact dimensions. This class of possible new gravitational instantons was previously considered in \cite{Gibbons:1979b}. In that paper, it was also argued that topological constraints, in the form of a Hitchin-type inequality, might rule out such new gravitational instantons for sufficiently large values of $|p|$. However, there are still some small values of $|p|\geq3$ for which new ALF or ALE gravitational instantons are not ruled out.

\newsubsection{Possible new gravitational instantons with three turning points}

It is straightforward to extend this analysis to find all possible rod structures with any given number of turning points, that would correspond to regular manifolds if appropriate identifications are made. In this subsection, we illustrate this for the next simplest case of three turning points. Again, we find that the allowed rod structures fall into two classes, depending on whether the two semi-infinite rods are parallel or not. The first class, with rod structure in standard orientation as shown in Fig.~5(a), has parallel semi-infinite rods. The infinity of the associated gravitational instanton will have topology $S^1\times S^2$, so it will likely be AF. The second class has rod structure in standard orientation as shown in Fig.~5(b), for any pair of integers $p$ and $q$ except $p=q=\pm1$. Asymptotically it will have a lens-space $L(pq-1,q)$ structure, with or without any compact dimensions.

Note that the triple-collinearly-centered Taub-NUT instanton (Fig.~3) emerges as a special case of the second class with $p=q=\pm2$. However, to the best of our knowledge, all the other cases would be associated to new gravitational instantons, if they exist. A systematic search is currently underway to construct new gravitational instantons with one or more of these allowed rod structures; indeed, preliminary results indicate that a new AF gravitational instanton with rod structure belonging to the first class (Fig.~5(a)) exists \cite{Chen2}. For the second class, topological or other constraints might serve to rule out certain ranges of values of $p$ and $q$. However, there might still exist new ALF or ALE gravitational instantons with sufficiently small values of $|p|$ and $|q|$.

\begin{figure}[t]
\begin{center}
\includegraphics{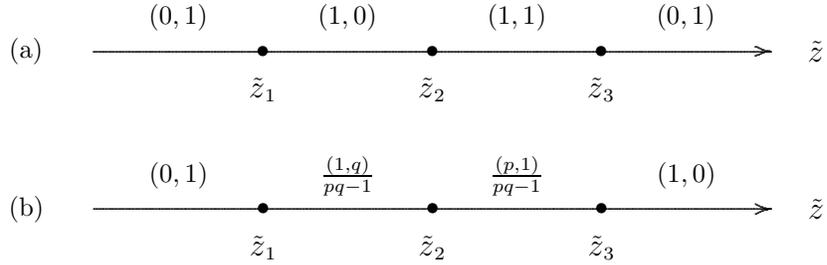}
\caption{The rod structure of possible new gravitational instantons with three turning points in standard orientation. In (b), $p$ and $q$ are any pair of integers except $p=q=\pm1$. Note that the special case $p=q=\pm2$ corresponds to the triple-collinearly-centered Taub-NUT instanton (Fig.~3).}
\end{center}
\end{figure}

\newsection{Discussion}

In this paper, we have defined a stronger version of the rod structure for five-dimensional black holes with $\mathbb{R}\times U(1)\times U(1)$ isometry, and explained how the regularity conditions can be read off from it. We then showed that the rod-structure formalism can also be usefully applied to the study of gravitational instantons with $U(1)\times U(1)$ isometry. A number of examples were considered, and several previous results concerning certain of these gravitational instantons were clarified in the process. It was then argued that the rod structure provides a way to classify all possible gravitational instantons with $U(1)\times U(1)$ isometry, and new classes of rod structures were explicitly written down. We finally speculated that at least some of these new rod structures would be associated to as yet undiscovered gravitational instantons.

It is interesting to further explore the (non-)uniqueness of the gravitational instantons analyzed in this paper. When restricted to the case when there is no black hole, Hollands and Yazadjiev's theorems \cite{Hollands:2007,Hollands:2008} immediately imply the following result: For gravitational instantons with $U(1)\times U(1)$ isometry, asymptotically approaching the Euclidean space $E^4$, or the product space $E^3\times S^1$ (with $S^1$ finite), there exists at most one gravitational instanton for a given rod structure. At infinity, the $U(1)\times U(1)$ isometry is assumed to generate the standard rotations of $E^4$, or of $E^3$ and $S^1$. It immediately follows that four-dimensional flat space and the Euclidean Schwarzschild instanton are the unique gravitational instantons that asymptotically approach $E^4$ and $E^3\times S^1$ respectively for their corresponding rod structures.

We have seen in Sec.~4 that, in certain cases, a gravitational instanton with $U(1)\times U(1)$ isometry cannot be uniquely determined by its rod structure. On the contrary, we even find that a one-parameter family of gravitational instantons can share the same rod structure. The simplest example is given by four-dimensional flat space and the self-dual Taub-NUT instanton, which share the same rod structure in standard orientation (as shown in Fig.~1). However, we notice that in this example the one-parameter degeneracy is resolved once the NUT charge of the self-dual Taub-NUT instanton, and thus its asymptotic geometry, is specified. Together with the result in the preceding paragraph, we may naturally conjecture that a gravitational instanton can be uniquely determined by its rod structure together with its asymptotic geometry, if specified in some appropriate way. Here, we also assume the various technical assumptions made in \cite{Hollands:2007,Hollands:2008} but with a vanishing black hole horizon holding in our case. One may further expect that this conjecture, if true, can be proved similarly as was done for the theorems in \cite{Hollands:2007,Hollands:2008}.

It would be interesting to know why the rod structure alone does not contain sufficient information to determine a gravitational instanton. Insights may be gained by studying the sources \cite{Harmark:2005}, and the relations between these sources and the corresponding rod structures of the gravitational instantons analyzed in this paper.

The open problem regarding the existence of a gravitational instanton for a given rod structure is even more challenging. Firstly, up to now not all the rod structures allowed by our analysis (so that conical and orbifold singularities can be removed) have been associated with a gravitational instanton, neither has their existence been disproved in general. For some of the new rod structures that we considered in Sec.~5, topological or other constraints might rule out the existence of any new associated gravitational instantons. If, however, these new gravitational instantons do exist, the inverse scattering method \cite{Belinski:2001,Pomeransky:2005} is a powerful solution-generating technique that might be able to construct them. Secondly, for a given rod structure, supposing their associated gravitational instantons exist, there seems to be constraints on the asymptotic geometry of these gravitational instantons. For example, gravitational instantons with a rod structure as shown in Fig.~2(c) (the Taub-bolt instanton) are only found to be ALF. An ALE gravitational instanton with such a rod structure cannot exist. This is because such a new gravitational instanton will have to be trivially AE (it has an infinity of topology $S^3$), and the positive action theorem \cite{Schoen:1979uj} rules out this possibility. On the other hand, gravitational instantons with a rod structure as shown in Fig.~2(b) can be either ALF (the double-centered Taub-NUT instanton) or ALE (the Eguchi--Hanson instanton).

Notice that a flat time dimension can be trivially added to the above gravitational instantons to obtain five-dimensional space-time solutions to the vacuum Einstein equations.\footnote{Such space-times, when constructed with AF or ALF gravitational instantons, have been studied in the context of Kaluza--Klein theory; see, e.g., \cite{Gross:1983} for details. When dimensionally reduced to four dimensions, they describe magnetic monopoles or dipoles, or their superpositions.} Moreover, static or stationary black holes may be added to such space-times, while preserving the $U(1)\times U(1)$ isometry. The black holes are then said to be sitting on that corresponding gravitational instanton, since when they are removed we recover a direct product of that gravitational instanton and a flat time dimension. Indeed, we have been able to classify or construct black holes on almost all the gravitational instantons studied in this paper with one or two turning points. These results will be presented in another paper \cite{Chen1}.

The rod-structure formalism developed in this paper may be readily generalised, in some modified form, to gravitational instantons with cosmological constant $\Lambda>0$. These instantons are compact Einstein manifolds \cite{Gibbons:1979c}. All the explicitly known gravitational instantons within this class, namely $S^4$, $\mathbb{C}P^2$, $S^2\times S^2$ and $\mathbb{C}P^2\,\sharp\, \overline{\mathbb{C}P^2}$ with their corresponding Einstein metrics  \cite{Gibbons:1979c,Perry:1980,Page:2009dm}, have the prescribed $U(1)\times U(1)$ isometry group. The concepts of turning points and rods are still applicable as fixed points of the $U(1)\times U(1)$ isometry group and its $U(1)$ isometry subgroups respectively. The direction of a rod is the normalised $2\pi$-periodic generator of the $U(1)$ isometry subgroup which (as a Killing vector field) vanishes along that rod. Two adjacent rods intersect at a turning point, with directions satisfying the constraint (\ref{condition for absence of orbifold singularity}). The direction pair of any two adjacent rods can then be identified as the pair of independent $2\pi$-periodic generators of the $U(1)\times U(1)$ isometry group. It turns out that the orbit spaces of these compact gravitational instantons are homeomorphic to a disk \cite{Orlik1,Orlik2}. The boundary, which is homeomorphic to a circle, is divided into arcs by the turning points. An arc on this circle is what was roughly referred to as a rod above. $S^4$, as an analytic continuation of the de Sitter space-time \cite{Gibbons:1977mu}, and $\mathbb{C}P^2$ \cite{Eguchi:1976db,Gibbons:1978zy} have two and three turning points, respectively. Both $S^2\times S^2$ \cite{Gibbons:1978zy} and $\mathbb{C}P^2\,\sharp\, \overline{\mathbb{C}P^2}$ \cite{Page:1979zv} have four turning points. The rod directions of these gravitational instantons can be easily calculated, and it turns out that the rod structures of $S^2\times S^2$ and $\mathbb{C}P^2\,\sharp\, \overline{\mathbb{C}P^2}$ can distinguish between these two gravitational instantons. We do not give any more details here.

Another possible generalization of the rod-structure formalism would be to consider $D$-dimensional black holes with $\mathbb{R}\times U(1)^{D-3}$ isometry and their background spaces, for $D>5$ \cite{Emparan:2001b,Harmark:2004,Hollands:2008,Chrusciel:2008}. The regular background spaces will be $(D-1)$-dimensional generalisations of gravitational instantons with $U(1)^{D-3}$ isometry. As already mentioned, necessary regularity conditions at the turning points of these space(-times) must be carefully treated \cite{Hollands:2008}. Higher-dimensional black holes and gravitational instantons with fewer isometries might also be treated by generalising the rod-structure formalism to the domain structure \cite{Harmark:2009}. For the case of five-dimensional stationary black holes, the possible horizon topologies have recently been classified in \cite{Hollands:2010}.

\bigskip\bigskip

{\renewcommand{\Large}{\normalsize}
}


\begin{thebibliography}{99}

\bibitem{Emparan:2001}
  R.~Emparan and H.~S.~Reall,
  ``A rotating black ring solution in five dimensions,''
  Phys.\ Rev.\ Lett.\  {\bf 88} (2002) 101101
  [arXiv:hep-th/0110260].

\bibitem{Myers:1986}
  R.~C.~Myers and M.~J.~Perry,
  ``Black holes in higher dimensional space-times,''
  Annals Phys.\  {\bf 172} (1986) 304.

\bibitem{Emparan:2008}
  R.~Emparan and H.~S.~Reall,
  ``Black holes in higher dimensions,''
  Living Rev.\ Rel.\  {\bf 11} (2008) 6
  [arXiv:0801.3471 [hep-th]].

\bibitem{Obers:2008}
  N.~A.~Obers,
  ``Black holes in higher-dimensional gravity,''
  Lect.\ Notes Phys.\  {\bf 769} (2009) 211
  [arXiv:0802.0519 [hep-th]].

\bibitem{Rodriguez:2010}
  M.~J.~Rodriguez,
  ``On the black hole species (by means of natural selection),''
  arXiv:1003.2411 [hep-th].

\bibitem{Emparan:2001b}
  R.~Emparan and H.~S.~Reall,
  ``Generalized Weyl solutions,''
  Phys.\ Rev.\  D {\bf 65} (2002) 084025
  [arXiv:hep-th/0110258].

\bibitem{Harmark:2004}
  T.~Harmark,
  ``Stationary and axisymmetric solutions of higher-dimensional general
  relativity,''
  Phys.\ Rev.\  D {\bf 70} (2004) 124002
  [arXiv:hep-th/0408141].

\bibitem{Harmark:2005}
  T.~Harmark and P.~Olesen,
  ``On the structure of stationary and axisymmetric metrics,''
  Phys.\ Rev.\  D {\bf 72} (2005) 124017
  [arXiv:hep-th/0508208].

\bibitem{Hollands:2007}
  S.~Hollands and S.~Yazadjiev,
  ``Uniqueness theorem for 5-dimensional black holes with two axial Killing
  fields,''
  Commum.\ Math.\ Phys.\  {\bf 283} (2008) 749
  [arXiv:0707.2775 [gr-qc]].

\bibitem{Hollands:2008}
  S.~Hollands and S.~Yazadjiev,
  ``A uniqueness theorem for stationary Kaluza--Klein black holes,''
  arXiv:0812.3036 [gr-qc].

\bibitem{Chrusciel:2008}
  P.~T.~Chrusciel,
  ``On higher dimensional black holes with abelian isometry group,''
  J.\ Math.\ Phys.\  {\bf 50} (2009) 052501
  [arXiv:0812.3424 [gr-qc]].

\bibitem{Chen1}
  Y.~Chen and E.~Teo,
  ``Black holes on gravitational instantons,''
  (in preparation).

\bibitem{Elvang:2005}
  H.~Elvang, R.~Emparan, D.~Mateos and H.~S.~Reall,
  ``Supersymmetric 4D rotating black holes from 5D black rings,''
  JHEP {\bf 0508} (2005) 042
  [arXiv:hep-th/0504125].

\bibitem{Gaiotto:2005}
  D.~Gaiotto, A.~Strominger and X.~Yin,
  ``5D black rings and 4D black holes,''
  JHEP {\bf 0602} (2006) 023
  [arXiv:hep-th/0504126].

\bibitem{Bena:2005}
  I.~Bena, P.~Kraus and N.~P.~Warner,
  ``Black rings in Taub-NUT,''
  Phys.\ Rev.\  D {\bf 72} (2005) 084019
  [arXiv:hep-th/0504142].

\bibitem{Newman:1963}
  E.~Newman, L.~Tamubrino and T.~Unti,
  ``Empty space generalization of the Schwarzschild metric,''
  J.\ Math.\ Phys.\  {\bf 4} (1963) 915.

\bibitem{Hawking:1976}
  S.~W.~Hawking,
  ``Gravitational instantons,''
  Phys.\ Lett.\  A {\bf 60} (1977) 81.

\bibitem{Gibbons:1979c}
  G.~W.~Gibbons and S.~W.~Hawking,
  ``Classification of gravitational instanton symmetries,''
  Commun.\ Math.\ Phys.\  {\bf 66} (1979) 291.

\bibitem{Gibbons:1994}
  G.~W.~Gibbons and S.~W.~Hawking,
  ``Euclidean quantum gravity,''
  World Scientific, Singapore (1993).

\bibitem{Gibbons:1976}
  G.~W.~Gibbons and S.~W.~Hawking,
  ``Action integrals and partition functions in quantum gravity,''
  Phys.\ Rev.\  D {\bf 15} (1977) 2752.

\bibitem{Eguchi:1978}
  T.~Eguchi and A.~J.~Hanson,
  ``Asymptotically flat selfdual solutions to Euclidean gravity,''
  Phys.\ Lett.\  B {\bf 74} (1978) 249.

\bibitem{Page:1979}
  D.~N.~Page,
  ``Taub-NUT instanton with an horizon,''
  Phys.\ Lett.\  B {\bf 78} (1978) 249.

\bibitem{Gibbons:1979b}
  G.~W.~Gibbons and M.~J.~Perry,
  ``New gravitational instantons and their interactions,''
  Phys.\ Rev.\  D {\bf 22} (1980) 313.

\bibitem{Gibbons:1979}
  G.~W.~Gibbons and S.~W.~Hawking,
  ``Gravitational multi-instantons,''
  Phys.\ Lett.\  B {\bf 78} (1978) 430.

\bibitem{Misner:1963}
  C.~W.~Misner,
  ``The flatter regions of Newman, Unti and Tamburino's generalized
  Schwarzschild space,''
  J.\ Math.\ Phys.\  {\bf 4} (1963) 924.

\bibitem{Ghezelbash:2007}
  A.~M.~Ghezelbash, R.~B.~Mann and R.~D.~Sorkin,
  ``The disjointed thermodynamics of rotating black holes with a NUT twist,''
  Nucl.\ Phys.\  B {\bf 775} (2007) 95
  [arXiv:hep-th/0703030].

\bibitem{Martelli:2005}
  D.~Martelli and J.~Sparks,
  ``Toric Sasaki-Einstein metrics on $S^2 \times S^3$,''
  Phys.\ Lett.\  B {\bf 621} (2005) 208
  [arXiv:hep-th/0505027].

\bibitem{Cvetic:2005}
  M.~Cvetic, H.~L\"u, D.~N.~Page and C.~N.~Pope,
  ``New Einstein-Sasaki and Einstein spaces from Kerr-de Sitter,''
  JHEP {\bf 0907} (2009) 082
  [arXiv:hep-th/0505223].

\bibitem{Lu:2008}
  H.~L\"u, J.~Mei and C.~N.~Pope,
  ``New black holes in five dimensions,''
  Nucl.\ Phys.\  B {\bf 806} (2009) 436
  [arXiv:0804.1152 [hep-th]].

\bibitem{Dowker:1995}
  F.~Dowker, J.~P.~Gauntlett, G.~W.~Gibbons and G.~T.~Horowitz,
  ``Nucleation of $p$-branes and fundamental strings,''
  Phys.\ Rev.\  D {\bf 53} (1996) 7115
  [arXiv:hep-th/9512154].

\bibitem{Orlik1}
  P.~Orlik and F.~Raymond,
  ``Actions of the torus on 4-manifolds I,''
  Transactions of the AMS {\bf 152} (1972) 531.

\bibitem{Orlik2}
  P.~Orlik and F.~Raymond,
  ``Actions of the torus on 4-manifolds II,''
  Topology {\bf 13} (1974) 89.

\bibitem{Eguchi:1978b}
  T.~Eguchi and A.~J.~Hanson,
  ``Selfdual solutions to Euclidean gravity,''
  Annals Phys.\  {\bf 120} (1979) 82.

\bibitem{Eguchi:1980}
  T.~Eguchi, P.~B.~Gilkey and A.~J.~Hanson,
  ``Gravitation, gauge theories and differential geometry,''
  Phys.\ Rept.\  {\bf 66} (1980) 213.

\bibitem{Perry:1980}
  M.~J.~Perry,
  ``Gravitational instantons,'' {\it in\/} S.-T.\ Yau, ed., ``Seminar on differential geometry,'' Princeton University Press, Princeton, N.J.\ (1982).

\bibitem{Gibbons:1979gd}
  G.~W.~Gibbons, C.~N.~Pope and H.~Romer,
  ``Index theorem boundary terms for gravitational instantons,''
  Nucl.\ Phys.\  B {\bf 157} (1979) 377.

\bibitem{Schoen:1979uj}
  R.~M.~Schoen and S.~T.~Yau,
  ``Proof of the positive action conjecture in quantum relativity,''
  Phys.\ Rev.\ Lett.\  {\bf 42} (1979) 547.

\bibitem{Feinblum}
  D.~A.~Feinblum,
  ``Global singularities and the Taub-NUT metric,''
  J.\ Math.\ Phys.\  {\bf 11} (1970) 2713.

\bibitem{Aharony:2002}
  O.~Aharony, M.~Fabinger, G.~T.~Horowitz and E.~Silverstein,
  ``Clean time-dependent string backgrounds from bubble baths,''
  JHEP {\bf 0207} (2002) 007
  [arXiv:hep-th/0204158].

\bibitem{Hunter:1998}
  C.~J.~Hunter,
  ``The action of instantons with nut charge,''
  Phys.\ Rev.\  D {\bf 59} (1999) 024009
  [arXiv:gr-qc/9807010].

\bibitem{Ida:2009}
  D.~Ida,
  ``On the topology of black lenses,''
  Prog.\ Theor.\ Phys.\  {\bf 122} (2010) 987
  [arXiv:0904.3581 [gr-qc]].

\bibitem{Chen2}
  Y.~Chen and E.~Teo,
  ``A new AF gravitational instanton,''
  (in preparation).

\bibitem{Belinski:2001}
  V.~Belinski and E.~Verdaguer,
  ``Gravitational solitons,''
  Cambridge University Press, U.K.\ (2001).

\bibitem{Pomeransky:2005}
  A.~A.~Pomeransky,
  ``Complete integrability of higher-dimensional Einstein equations with
  additional symmetry, and rotating black holes,''
  Phys.\ Rev.\  D {\bf 73} (2006) 044004
  [arXiv:hep-th/0507250].

\bibitem{Gross:1983}
  D.~J.~Gross and M.~J.~Perry,
  ``Magnetic monopoles in Kaluza--Klein theories,''
  Nucl.\ Phys.\  B {\bf 226} (1983) 29.

\bibitem{Page:2009dm}
  D.~N.~Page,
  ``Some gravitational instantons,''
  arXiv:0912.4922 [gr-qc].

\bibitem{Gibbons:1977mu}
  G.~W.~Gibbons and S.~W.~Hawking,
  ``Cosmological event horizons, thermodynamics, and particle creation,''
  Phys.\ Rev.\  D {\bf 15} (1977) 2738.

\bibitem{Eguchi:1976db}
  T.~Eguchi and P.~G.~O.~Freund,
  ``Quantum gravity and world topology,''
  Phys.\ Rev.\ Lett.\  {\bf 37} (1976) 1251.

\bibitem{Gibbons:1978zy}
  G.~W.~Gibbons and C.~N.~Pope,
  ``$\mathbb{C}P^2$ as a gravitational instanton,''
  Commun.\ Math.\ Phys.\  {\bf 61} (1978) 239.

\bibitem{Page:1979zv}
  D.~N.~Page,
  ``A compact rotating gravitational instanton,''
  Phys.\ Lett.\  B {\bf 79} (1978) 235.

\bibitem{Harmark:2009}
  T.~Harmark,
  ``Domain structure of black hole space-times,''
  Phys.\ Rev.\  D {\bf 80} (2009) 024019
  [arXiv:0904.4246 [hep-th]].

\bibitem{Hollands:2010}
  S.~Hollands, J.~Holland and A.~Ishibashi,
  ``Further restrictions on the topology of stationary black holes in five
  dimensions,''
  arXiv:1002.0490 [gr-qc].


\end{thebibliography}
\end{document}